\documentclass[twocolumn]{aastex631}

\newcommand\nustar{\textit{NuSTAR}}
\newcommand\nicer{\textit{NICER}}
\newcommand{\swift}{\textit{Swift}}
\newcommand{\srgt}{SRGA\,J181414.6-225604}
\newcommand{\artxc}{Mikhail Pavlinsky ART-XC}

\shorttitle{A Galactic Symbiotic X-ray Binary triggered by a dust formation episode}
\shortauthors{K. De et al.}

\graphicspath{{./}{figures/}}

\begin{document}

\title{SRGA\,J181414.6-225604: A new Galactic symbiotic X-ray binary outburst triggered by an intense mass loss episode of a heavily obscured Mira variable}

\author[0000-0002-0786-7307]{Kishalay De}
\altaffiliation{NASA Einstein Fellow}
\affiliation{MIT-Kavli Institute for Astrophysics and Space Research, 77 Massachusetts Ave., Cambridge, MA 02139, USA}

\author{Ilya Mereminskiy}
\affiliation{Space Research Institute, Russian Academy of Sciences, Profsoyuznaya 84/32, 117997 Moscow, Russia}

\author{Roberto Soria}
\affiliation{College of Astronomy and Space Sciences, University of the Chinese Academy of Sciences, Beijing 100049, China}
\affiliation{Sydney Institute for Astronomy, School of Physics A28, The University of Sydney, Sydney, NSW 2006, Australia}

\author{Charlie Conroy}
\affil{Harvard-Smithsonian Center for Astrophysics, Cambridge, MA 02138, USA}

\author{Erin Kara}
\affiliation{MIT-Kavli Institute for Astrophysics and Space Research, 77 Massachusetts Ave., Cambridge, MA 02139, USA}

\author{Shreya Anand}
\affil{Cahill Center for Astrophysics, California Institute of Technology, 1200 E. California Blvd. Pasadena, CA 91125, USA}

\author[0000-0003-1412-2028]{Michael C. B. Ashley}
\affil{School of Physics, University of New South Wales, Sydney NSW 2052, Australia}

\author{Martha L. Boyer}
\affil{Space Telescope Science Institute, 3700 San Martin Drive, Baltimore, MD 21218, USA}

\author{Deepto Chakrabarty}
\affiliation{MIT-Kavli Institute for Astrophysics and Space Research, 77 Massachusetts Ave., Cambridge, MA 02139, USA}

\author{Brian Grefenstette}
\affil{Cahill Center for Astrophysics, California Institute of Technology, 1200 E. California Blvd. Pasadena, CA 91125, USA}

\author[0000-0001-9315-8437]{Matthew J. Hankins}
\affil{Arkansas Tech University, Russellville, AR 72801, USA}

\author{Lynne A. Hillenbrand}
\affil{Cahill Center for Astrophysics, California Institute of Technology, 1200 E. California Blvd. Pasadena, CA 91125, USA}

\author[0000-0001-5754-4007]{Jacob E. Jencson}
\affil{Steward Observatory, University of Arizona, 933 North Cherry Avenue, Tucson, AZ85721-0065, USA}

\author{Viraj Karambelkar}
\affil{Cahill Center for Astrophysics, California Institute of Technology, 1200 E. California Blvd. Pasadena, CA 91125, USA}

\author[0000-0002-5619-4938]{Mansi M. Kasliwal}
\affil{Cahill Center for Astrophysics, California Institute of Technology, 1200 E. California Blvd. Pasadena, CA 91125, USA}

\author{Ryan M. Lau}
\affil{Institute of Space \& Astronautical Science, Japan Aerospace Exploration  Agency,  3-1-1 Yoshinodai,  Chuo-ku,  Sagamihara, Kanagawa 252-5210, Japan}

\author{Alexander Lutovinov}
\affiliation{Space Research Institute, Russian Academy of Sciences, Profsoyuznaya 84/32, 117997 Moscow, Russia}

\author{Anna M. Moore}
\affil{Research School of Astronomy and Astrophysics, Australian National University, Canberra, ACT 2611, Australia}

\author{Mason Ng}
\affiliation{MIT-Kavli Institute for Astrophysics and Space Research, 77 Massachusetts Ave., Cambridge, MA 02139, USA}

\author{Christos Panagiotou}
\affiliation{MIT-Kavli Institute for Astrophysics and Space Research, 77 Massachusetts Ave., Cambridge, MA 02139, USA}

\author{Dheeraj R. Pasham}
\affiliation{MIT-Kavli Institute for Astrophysics and Space Research, 77 Massachusetts Ave., Cambridge, MA 02139, USA}

\author{Andrey Semena}
\affiliation{Space Research Institute, Russian Academy of Sciences, Profsoyuznaya 84/32, 117997 Moscow, Russia}

\author{Robert Simcoe}
\affiliation{MIT-Kavli Institute for Astrophysics and Space Research, 77 Massachusetts Ave., Cambridge, MA 02139, USA}

\author{Jamie Soon}
\affil{Research School of Astronomy and Astrophysics, Australian National University, Canberra, ACT 2611, Australia}

\author{Gokul P. Srinivasaragavan}
\affil{Department of Astronomy, University of Maryland, College Park, MD 20742, USA.}

\author[0000-0001-9304-6718]{Tony Travouillon}
\affil{Research School of Astronomy and Astrophysics, Australian National University, Canberra, ACT 2611, Australia}

\author{Yuhan Yao}
\affil{Cahill Center for Astrophysics, California Institute of Technology, 1200 E. California Blvd. Pasadena, CA 91125, USA}

\correspondingauthor{Kishalay De}
\email{kde1@mit.edu}

\begin{abstract}

We present the discovery and multi-wavelength characterization of \srgt, a Galactic hard X-ray transient discovered during the ongoing SRG/ART-XC sky survey. Using data from the Palomar Gattini-IR survey, we identify a spatially and temporally coincident variable infrared (IR) source, IRAS\,18111-2257, and classify it as a very late-type (M7-M8), long period ($1502 \pm 24$\,days) and luminous ($M_K\approx -9.9 \pm 0.2$) O-rich Mira donor star located at a distance of $\approx 14.6^{+2.9}_{-2.3}$\,kpc. Combining multi-color photometric data over the last $\approx 25$\,years, we show that the IR counterpart underwent a recent (starting $\approx 800$\,days before the X-ray flare) enhanced mass loss (reaching $\approx 2.1 \times 10^{-5}$\,M$_\odot$\,yr$^{-1}$) episode resulting in an expanding dust shell obscuring the underlying star. Multi-epoch follow-up from \swift, \nicer\ and \nustar\ reveal a $\approx 200$\,day long X-ray outburst reaching a peak luminosity of $L_X \approx 2.5 \times 10^{36}$\,erg\,s$^{-1}$, characterized by a heavily absorbed ($N_{\rm H} \approx 6\times 10^{22}$\,cm$^{-2}$) X-ray spectrum consistent with an optically thick Comptonized plasma. The X-ray spectral and timing behavior suggest the presence of clumpy wind accretion together with a dense ionized nebula overabundant in silicate material surrounding the compact object. Together, we show that \srgt\ is a new symbiotic X-ray binary in outburst, triggered by an intense dust formation episode of a highly evolved donor. Our results offer the first direct confirmation for the speculated connection between enhanced late-stage donor mass loss and active lifetimes of the symbiotic X-ray binaries.
\end{abstract}


\section{Introduction} \label{sec:intro}

Symbiotic binaries consist of a compact object (white dwarf, neutron star or black hole) accreting from the wind of a late-type giant companion \citep{Kenyon1986, Belczynski2000}. They are broadly classified into two groups based on their near-infrared (NIR) spectral energy distributions (SEDs): i) the S-type systems exhibiting SEDs characteristic of a cool photosphere with temperature $T_{\rm eff} \approx 3000-4000$\,K and ii) the D-type systems with SEDs characteristic of warm dust with $T_{\rm eff}\approx 1000$\,K \citep{Allen1974, Webster1975}. The recent compilation of \citet{Akras2019} suggests that only $\approx 15$\% of known symbiotic stars are identified with D-type SEDs. Spectral classification of symbiotics \citep{Murset1999}, together with light curves from photometric surveys \citep{Whitelock1987, Gromadzki2009} suggest that D-type systems contain late type Mira variables as donor stars \citep{Phillips2007, Angeloni2010}. Mira symbiotics are known to exhibit long pulsation periods \citep{Whitelock1987} and thick dust shells that give rise to their dust dominated SEDs \citep{Muerset1996, Angeloni2010}.

Symbiotic X-ray binaries (SyXRBs; \citealt{Masetti2006}, see \citealt{Kuranov2015} for a review) form a sub-class of the symbiotic binary population, exhibiting X-ray emission extending up to high energies ($E\gtrsim 2.4$\,keV) produced by optically thick Comptonizing plasma \citep{Masetti2007, Luna2013, Lutovinov2020} from accretion on to a neutron star (NS) or black hole (BH). SyXRBs are relatively rare among Galactic XRBs, with only about a dozen systems known \citep{Yungelson2019}, all of them believed to contain NSs. The accretion of the companion wind onto the compact object produces relatively low average X-ray luminosity ($L_X \sim 10^{32-36}$\,erg\,s$^{-1}$), rendering them difficult to detect in shallow X-ray surveys. Unlike their white dwarf (WD) counterparts, the survival of a wide orbit companion during the supernova (SN) that created the compact object requires fine-tuned mass loss and kick parameters \citep{Iben1995}, making SyXRBs unique probes of extremes in binary evolution physics.

Binary population synthesis models predict that there are $\approx 50 - 1000$ SyXRBs in the Galaxy \citep{Lu2012, Kuranov2015, Yungelson2019} compared to roughly dozen known candidate and confirmed systems.  It is generally understood that SyXRBs become conspicuous in the X-ray sky only for a brief fraction of the binary lifetime when the donor produces a dense wind, a fraction of which is captured by the compact object  \citep{Hinkle2006}. The most well-known SyXRB GX\,1+4 ($L_X \sim 10^{37-38}$\,erg\,s$^{-1}$) indeed has a very late type (M6-III) donor at the tip of the Red Giant Branch \citep{Chakrabarty1997}. While the first SyXRBs were identified as persistent and highly variable X-ray sources, repeated scans of the hard X-ray sky from all-sky surveys are offering new avenues to finding outbursting sources where the persistent emission is too faint to be detected. Although an understanding of these transient flares (lasting $\sim$weeks to $\sim$years) is crucial to establishing their active lifetimes and Galactic demographics, their nature remains poorly constrained.

Previous speculations attribute the accretion variability to the motion of the compact object in an eccentric orbit \citep{Pereira1999, Masetti2002, Ikiewicz2017}, to changes in the donor star's mass loss properties \citep{Masetti2002, Galloway2002, Bozzo2018} or to changes in the angular momentum of the accreted material near the NS magnetosphere \citep{Postnov2010, Marcu2011, Shakura2012}. While temporal characterization of the donor stars offers a promising pathway to directly test these scenarios, the intrinsically red donor SEDs together with heavy dust reddening preclude detailed optical investigations of the donor variability and its impact on the accretion rate on to the compact object. The timely emergence of deep all-sky hard X-ray surveys together with near-infrared (NIR) time domain surveys offers a new opportunity to understand this population. 

In this paper, we present the discovery of \srgt, a hard X-ray transient discovered by the \artxc\ instrument on-board the SRG satellite. Combining archival photometry from time domain surveys and extensive follow-up in the X-ray, optical and infrared bands, we establish the source as a new D-type Galactic SyXRB triggered by a recent intense mass loss episode of a highly evolved Mira-like variable. Section \ref{sec:obs} presents details of the discovery, counterpart identification and follow-up observations. Section \ref{sec:xrayprop} presents an analysis of the spectral and temporal behavior of the X-ray transient. In Section \ref{sec:ircounterpart}, we constrain the nature of the donor star with optical/infrared photometry and spectroscopy, while we present evidence for a recent massive dust obscuration episode in Section \ref{sec:dustobs}. We end with a discussion on the nature of the binary system in Section \ref{sec:discussion} and summarize our findings in Section \ref{sec:summary}.

\section{Observations and Data Reduction}
\label{sec:obs}
\subsection{ART-XC Discovery and Infrared identification}

The Spektr-RG (SRG) mission \citep{Sunyeav2021}, equipped with the \artxc\ \citep{Pavlinsky2021} and eROSITA \citep{Predehl2020} instruments, is carrying out the deepest all-sky X-ray survey, re-visiting the entire sky every $\approx 6$\,months for a total duration of $\approx 4$\,years. On UT 2021-04-01, the ART-XC instrument discovered a bright X-ray source at J2000 coordinates $\alpha =$ 18:14:14, $\delta =$ -22:56:04 with a 4-12\,keV X-ray flux of $\approx 4 \times 10^{-11}$\,erg\,cm$^{-2}$\,s$^{-1}$ \citep{Mereminskiy2021}. The source was localized to a $90$\% error radius of $\approx 15$\arcsec. The source was not detected in the two previous ART-XC surveys individually to a depth of $\approx 8\times 10^{-12}$\,erg\,cm$^{-2}$\,s$^{-1}$ in the 4 - 12 keV band, but marginally detected by combining the two epochs (see Section \ref{sec:xray_arc}). The source was also detected by the eROSITA instrument with an absorbed and hard X-ray spectrum at a flux level of $\approx 6.4 \times 10^{-12}$\,erg\,cm$^{-2}$\,s$^{-1}$ in the 0.5 - 6 keV band. The eROSITA instrument provided an improved localization of $\alpha =$ 18:14:15.1, $\delta = $  -22:56:17.0 with a 95\% error radius of $\approx 9$\arcsec.

\begin{figure}[!ht]
    \centering
    \includegraphics[width=\columnwidth]{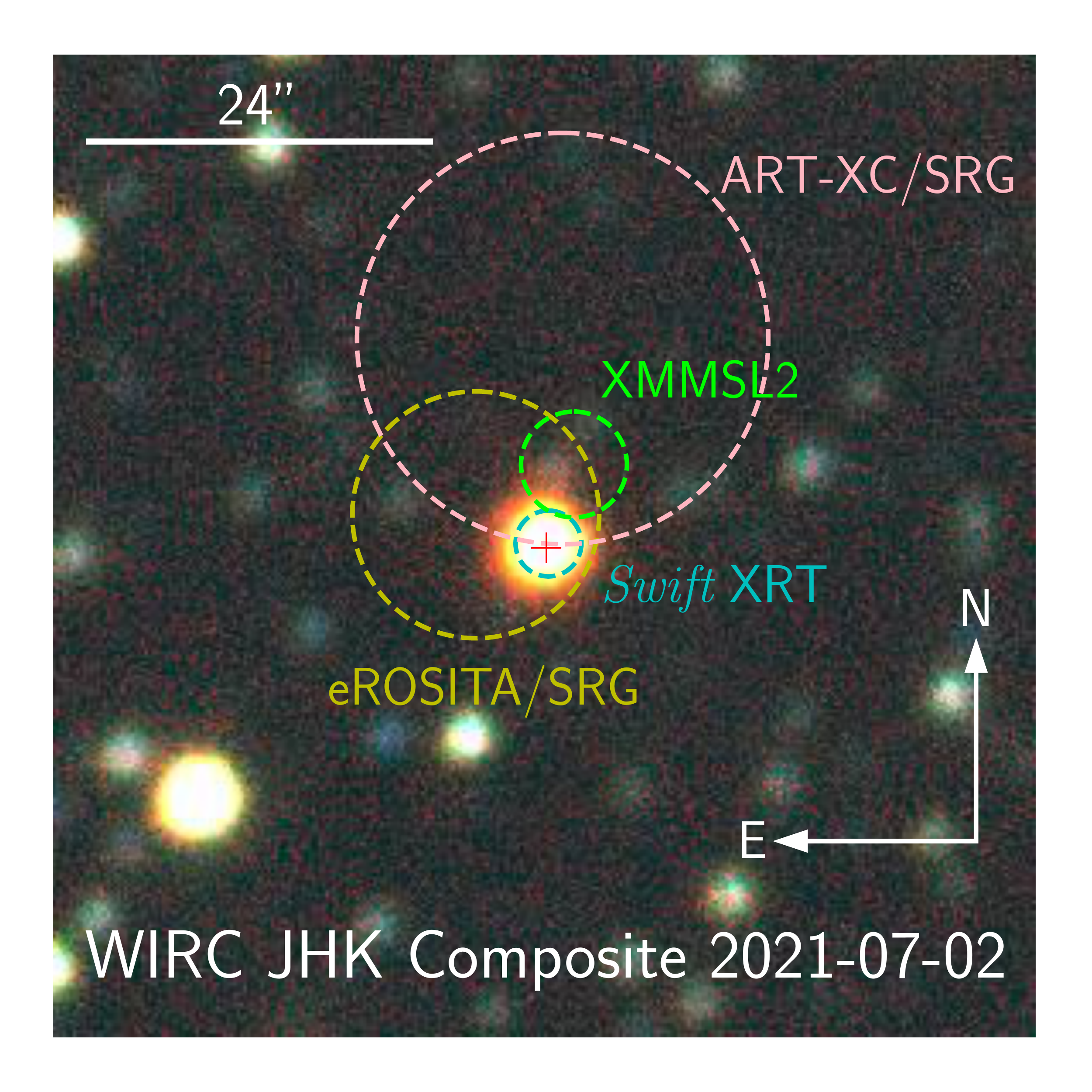}
    \caption{NIR composite false-color image of the field of IRAS\,18111-2257 obtained with WIRC on P200. The image shows the 90\% confidence localization regions for \srgt\ reported by ART-XC and eROSITA at the time of discovery, as well as the position obtained from the \swift\ XRT follow-up campaign. The image is centered on IRAS\,18111-2257 (shown with a red cross), which is consistent within $0.4$\arcsec\ of the best-fit location from the \swift\ XRT observations. The green circle shows the $1\,\sigma$ localization region of a coincident X-ray source reported in the {\it XMM Newton} Slew Survey catalog (see text). }
    \label{fig:localization}
\end{figure}

Following the discovery of the X-ray transient, we identified a highly variable infrared counterpart within the eROSITA localization region at a magnitude of $J \approx 10.2 - 12.5$\,mag in data from the Palomar Gattini-IR (PGIR) NIR time domain survey \citep{Moore2019, De2020a}. The variable NIR source (also known as 2MASS\,J18141475-2256195) was spatially coincident with the bright mid-IR source IRAS\,18111-2257 listed in the SIMBAD database\footnote{\url{https://simbad.u-strasbg.fr/simbad/sim-id?Ident=IRAS+18111-2257}}. The PGIR data revealed a long-term NIR brightening coincident with the X-ray flare, suggesting a likely association \citep{De2021}. The association was subsequently supported by a follow-up observation with the {\it Neil Gehrels Swift} Observatory ({\it Swift}), which localized the X-ray transient to within 0.8\arcsec\ of the IR source \citep{Heinke2021}.

On UT 2021-04-07, we obtained a very low resolution optical spectrum of the source with the SED Machine Spectrograph \citep{Blagorodnova2018} on the Palomar 60-inch telescope. The spectrum was reduced with the \texttt{pysedm} pipeline \citep{Rigault2019}, and exhibited a steep red continuum with broad absorption features consistent with VO and TiO absorption in late-type stars, suggesting that the source was a likely new Galactic symbiotic X-ray binary in outburst. Figure \ref{fig:localization} shows a NIR image of the localization region of the X-ray transient, together with best derived positions from the X-ray discovery and follow-up observations.

\subsection{Optical/IR Photometry}

\begin{table}[]
    \footnotesize
    \centering
    \caption{{Ground-based photometry of IRAS\,18111-2257, the infrared counterpart of \srgt. Upper limits are denoted at $3\sigma$ confidence. The entire table will be available electronically upon publication.}}
    \begin{tabular}{ccc}
    \hline
    \hline
       MJD  & Flux & Instrument/Filter  \\
         &  Mag & \\
        \hline
$58011.29$ & $16.46 \pm 0.01$ & ATLAS/$o$\\ 
$58019.28$ & $16.34 \pm 0.03$ & ATLAS/$o$\\ 
$58031.73$ & $16.48 \pm 0.01$ & ATLAS/$o$\\ 
$58037.71$ & $16.41 \pm 0.01$ & ATLAS/$o$\\ 
$58047.89$ & $16.39 \pm 0.01$ & ATLAS/$o$\\ 
$58057.49$ & $16.41 \pm 0.02$ & ATLAS/$o$\\ 
$58066.53$ & $16.40 \pm 0.02$ & ATLAS/$o$\\ 
$58158.67$ & $16.33 \pm 0.09$ & ATLAS/$o$\\ 
$58183.22$ & $16.43 \pm 0.02$ & ATLAS/$o$\\ 
$58207.51$ & $16.38 \pm 0.02$ & ATLAS/$o$\\ 
\hline
    \end{tabular}
    \label{tab:oirlc}
\end{table}

\begin{figure*}[!ht]
    \centering
    \includegraphics[width=\textwidth]{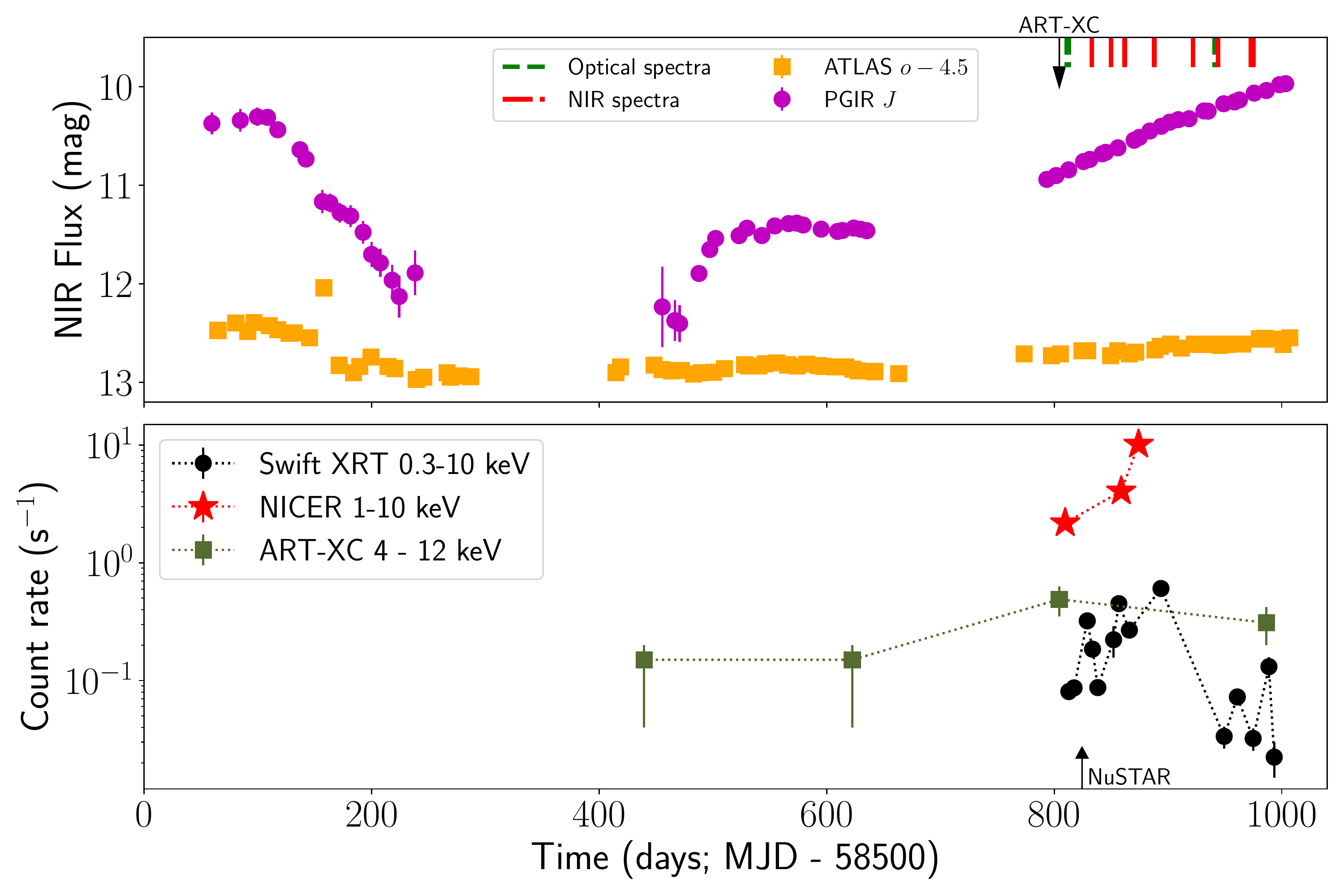}
    \caption{Multi-wavelength light curves of \srgt\ during 2019-2021. {\it Top panel}: Near-infrared photometric evolution of the identified counterpart from the PGIR and ATLAS surveys. The black arrow shows the epoch of first identification of the transient from ART-XC, co-incident with the onset of a NIR brightening. Epochs of optical/NIR spectroscopy are also shown. {\it Bottom panel}: The X-ray count rate evolution of \srgt\ in data from the SRG survey as well as follow-up from \swift\ XRT and \nicer. The NuSTAR observation epoch is shown with a black arrow on the lower axis.}

    \label{fig:srg_lc}
\end{figure*}

The highly reddened NIR counterpart of \srgt, IRAS\,18111-2257, was monitored as a part of regular survey operations of PGIR, and we derived a $J$-band light curve of the source as described in Appendix \ref{sec:app_photo}. In addition, we searched all publicly available data from time domain surveys to derive multi-color light curves of the counterpart. As described in Appendix \ref{sec:app_photo}, the variable source is detected in the $o$-filter of the Asteroid Terrestrial-impact Last Alert System (ATLAS; \citealt{Tonry2018}) survey, the $ri$ filters of the Zwicky Transient Facility (ZTF; \citealt{Bellm2019}) survey, the $grizy$ filters of the PanSTARRS (PS1; \citealt{Chambers2016}) survey and in the Wide-field Infared Survey Explorer surveys \citep{Wright2010,Mainzer2011}. Figure \ref{fig:srg_lc} shows the optical/IR photometric evolution of the counterpart during 2019-2021, in data from PGIR and ATLAS which have the highest cadence observations around the X-ray brightening of the source.

\begin{table}[!ht]
\footnotesize
    \centering
    \caption{Multi-epoch photometry of IRAS\,18111-2257 from space-based instruments. The entire table will be available electronically upon publication.}
    \begin{tabular}{ccc}
    \hline
    \hline
    MJD &  Flux & Instrument/Filter \\
     & Mag & \\
    \hline
$55279.01$ & $5.46 \pm 0.15$ & WISE/W1\\
$55461.39$ & $5.72 \pm 0.15$ & WISE/W1\\
$56743.67$ & $5.54 \pm 0.24$ & WISE/W1\\
$56923.43$ & $6.30 \pm 0.16$ & WISE/W1\\
$57105.59$ & $6.31 \pm 0.19$ & WISE/W1\\
$57282.40$ & $5.92 \pm 0.17$ & WISE/W1\\
$57469.96$ & $5.84 \pm 0.18$ & WISE/W1\\
$57644.34$ & $5.52 \pm 0.21$ & WISE/W1\\
$57837.06$ & $5.33 \pm 0.22$ & WISE/W1\\
$58004.57$ & $5.65 \pm 0.19$ & WISE/W1\\
\hline
    \end{tabular}
    \label{tab:space_photo}
\end{table}

On UT 2021-04-09, we obtained additional $gri$ follow-up photometry with the rainbow camera on the Palomar 60-inch telescope. The source is only detected in the $i$ filter. On UT 2021-07-02, we obtained multi-color NIR photometry of the source in the $J$, $H$ and $Ks$ band using the Wide-field Infrared Camera (WIRC; \citealt{Wilson2003}) on the Palomar 200-inch telescope. We obtained a series of dithered exposures of $\approx 0.9$\,s each, which were reduced, stacked and photometrically calibrated using the pipeline described in \citet{De2020a}. Figure \ref{fig:localization} shows the multi-color composite NIR image of the source obtained from the WIRC data, and Table \ref{tab:oirlc} provides the complete ground-based photometric dataset. Table \ref{tab:space_photo} provides the photometric dataset derived from space-based archival/follow-up observations.

The source was also observed by several surveys prior to 2010, including the original identification of the bright mid-infrared source by the Infrared Astronomical Satellite (IRAS; \citealt{Neugebauer1984, Abrahamyan2015}), the Midcourse Space Experiment (MSX; \citealt{Egan2003}), the Two Micron All-sky Survey (2MASS; \citealt{Skrutskie2006}), the Deep Near-Infrared Survey of the Southern Sky (DENIS\footnote{\url{http://cdsweb.u-strasbg.fr/denis.html}}), AKARI \citep{Murakami2007} and the Spitzer Galactic Legacy Infrared Mid-Plane Survey (GLIMPSE; \citealt{Benjamin2003}). In Table \ref{tab:archival_photo}, we provide the reported archival fluxes of the source along with the approximate date of observation.

\begin{table}[]
\footnotesize
    \caption{Archival photometry of IRAS\,18111-2257. In cases where published magnitudes were available, magnitudes were converted to fluxes in Jansky using published zero-point fluxes for the respective surveys.}
    \centering
    \begin{tabular}{cccc}
    \hline
    \hline
    Survey & Year & Band & Flux (Jy)\\
    \hline
    IRAS & $\approx 1983$ & $12\,\mu$m & $1.77$\\
    IRAS & $\approx 1983$ & $25\,\mu$m & $0.68$\\
    IRAS & $\approx 1983$ & $60\,\mu$m & $< 7.44$\\
    IRAS & $\approx 1983$ & $100\,\mu$m & $< 106$\\
    MSX\_A & $\approx 1996$ & $8.28\,\mu$m & $0.98 \pm 0.05$\\
    MSX\_C & $\approx 1996$ & $12.1\,\mu$m & $1.21 \pm 0.09$\\
    MSX\_D & $\approx 1996$ & $14.6\,\mu$m & $0.80 \pm 0.07$\\
    2MASS & $\approx 1998$ & $J$ & $0.74 \pm 0.02$ \\
    2MASS & $\approx 1998$ & $H$ &  $1.73 \pm 0.06$ \\
    2MASS & $\approx 1998$ & $K$ & $2.30 \pm 0.04$ \\
    DENIS & $\approx 1998$ &  $J$ & $0.49 \pm 0.04$\\
    DENIS & $\approx 1998$ &  $K$ & $5.59 \pm 1.29$ \\    
    AKARI & $\approx 2007$ & $L18$ & $0.56 \pm 0.02$ \\
    GLIMPSE & $\approx 2006$ & $4.5\mu$m & $1.03 \pm 0.05$\\
    GLIMPSE & $\approx 2006$ & $8.0\mu$m & $0.77 \pm 0.02$\\
    \hline
    \end{tabular}
    \label{tab:archival_photo}
\end{table}

\subsection{Optical/IR spectroscopy}

\begin{table}[!ht]
\footnotesize
    \caption{Spectroscopic follow-up of \srgt\ in the optical and NIR bands}
    \centering
    \begin{tabular}{cccc}
    \hline
    \hline
         UT Date & Instrument & $\lambda$ Range & $R$  \\
          & & $\mu$m & \\
          \hline
         2021-04-08 & ANU\,2.3m + WiFeS & 0.4 - 0.95 & 3000\\
         2021-04-09 & P200 + DBSP & 0.36 - 1.00 & 1000 \\
         2021-04-29 & P200 + TSpec & 1.00 - 2.46 & 2600 \\
         2021-05-16 & IRTF + SpeX & 0.75 - 2.55 & 2000 \\
         2021-05-16 & IRTF + SpeX & 2.00 - 5.35 & 2000 \\
         2021-05-28 & P200 + TSpec & 1.00 - 2.46 & 2600 \\
         2021-06-23 & IRTF + SpeX & 0.75 - 2.55 & 2000 \\
         2021-07-27 & P200 + TSpec & 1.00 - 2.46 & 2600 \\
         2021-08-15 & P200 + DBSP & 0.36 - 1.00 & 1000 \\
         2021-08-18 & Magellan + FIRE & 0.85 - 2.49 & 5000\\
         2021-09-16 & P200 + TSpec & 1.00 - 2.46 & 2600 \\
         2021-09-18 & Magellan + FIRE & 0.85 - 2.49 & 5000\\
         \hline
    \end{tabular}
    \label{tab:spec}
\end{table}

We obtained three epochs of follow-up optical spectroscopy of the source using the Wide-field Spectrograph (WiFeS; \citealt{Dopita2010}) on the ANU 2.3 m telescope at Siding Spring Observatory and the Double Spectrograph (DBSP; \citealt{Oke1982}) on the Palomar 200-inch telescope. The WiFeS data were reduced using the PyWiFeS pipeline \citep{Childress2014} and the DBSP data were reduced using the \texttt{pyraf-dbsp} pipeline \citep{Bellm2016b}. We obtained a total of eight epochs of NIR spectroscopy using the TripleSpec spectrograph on the Palomar 200-inch telescope \citep{Herter2008}, SpeX on the NASA Infrared Telescope Facility (\citealt{Rayner2003}; Program 2021A083, PI: De) and the Folded-port Infrared Echellete (FIRE; \citealt{Simcoe2008, Simcoe2010}) spectrograph on the Magellan 6.5 m telescope. The TripleSpec and SpeX data were reduced using the \texttt{spextool} pipeline \citep{Cushing2004}, while telluric and flux calibration were performed using the \texttt{xtellcor} package \citep{Vacca2003}. The FIRE data were reduced, flux calibrated and corrected for telluric absorption using the \texttt{pypeit} package \citep{pypeit}. Table \ref{tab:spec} summarizes all the epochs of spectroscopy of the source, while they are shown by vertical lines in Figure \ref{fig:srg_lc} along the photometric evolution.

\subsection{X-ray coverage from surveys}
\label{sec:xray_arc}
\begin{table*}[!ht]
\footnotesize
    \caption{Long-term X-ray flux constraints on \srgt\ from wide-area surveys. Count rates were converted to observed X-ray fluxes in the same energy range assuming a power law spectrum with photon index $\Gamma = 1.5$ and column density of $N_{\rm H} = 6 \times 10^{22}$\,cm$^{-2}$.}
    \centering
    \begin{tabular}{ccccc}
    \hline
    \hline
    Mission & Date & Energy Band & Count Rate & Observed Flux \\
     & MJD & keV & ct\,s$^{-1}$ & $10^{-11}$\,erg\,cm$^{-2}$\,s$^{-1}$\\
    \hline
    {\it ROSAT} & 49240.53 & 0.2 - 2 & $< 0.04$ & $<0.11$\\
    {\it INTEGRAL} & 52640 - 58850 & 17 - 60 & - & $< 0.3$\\
    {\it XMM} SL & 52921.62 & 0.2 - 12 & $< 1.22$ & $<1.57$\\
    & 55295.28 & 0.2 - 12 & $0.87^{+0.42}_{-0.42}$ & $1.12^{+0.54}_{-0.54}$\\
    & 	56730.72 & 0.2 - 12 & $< 0.89$ & $< 1.15$\\
    & 57279.50 & 0.2 - 12 & $1.66^{+0.56}_{-0.56}$ & $2.14^{+0.72}_{-0.72}$\\
    SRG/ART-XC & 58943-59130   &   4-12   &  $0.15^{+0.05}_{-0.10}$ & $0.6_{-0.3}^{+0.2}$\\
          &  59305-59306    &   4-12   &  $0.49^{+0.14}_{-0.14}$ & $2.2^{+0.6}_{-0.6}$\\
          &  59489-59490    &   4-12   &  $0.31^{+0.11}_{-0.11}$ & $1.4_{-0.5}^{+0.5}$\\
    \hline
    \end{tabular}

    \label{tab:xray_archive}
\end{table*}

As noted in \citet{Mereminskiy2021}, an X-ray source is reported in the {\it XMM Newton} Slew Survey catalog \citep{Saxton2008} within the ART-XC localization, as XMMSL2\,J181414.6-225613. The reported location of the {\it XMM} Slew Survey source along with its uncertainty is shown in the NIR image in Figure \ref{fig:localization}. The spatial coincidence (within  $\approx 2 \sigma \approx 7$\,\arcsec) suggests that XMMSL2\,J181414.6-225613 is likely associated with \srgt. To derive the complete history of X-ray data at this position, we used the High Energy Light Curve Generator\footnote{\url{http://xmmuls.esac.esa.int/upperlimitserver/}}\citep{Saxton2021} at the location of the IR counterpart. The source position was covered over four epochs in the {\it XMM} Slew Surveys, including two epochs of X-ray detection in 2010 and 2015 respectively. The source was also covered by the ROSAT \citep{Voges1999} and INTEGRAL sky surveys \citep{Krivonos2021} but not detected.  We summarize the archival X-ray coverage in Table \ref{tab:xray_archive}.

During the all-sky survey, ART-XC covers the same sky position nearly each six months. We reprocessed all data, covering the \srgt\, position using the ART-XC pipeline {\sc artproducts} v0.9 and {\sc CALDB} version 20200401. The source is clearly detected in the 4-12 keV energy band during third and fourth surveys (April, 01-02 and October, 02-03, 2021). Although initially undetected, forced photometry of the stacked data from the first two surveys also provides a low significance ($\approx$3\,$\sigma$) detection of the source at a flux level of $\approx 6_{-3}^{+2} \times 10^{-12}$\,erg\,cm$^{-2}$\,s$^{-1}$. We are unable to assess the source variability during the first two surveys due to its faint flux. Due to the small number photons detected from the source, we adopt a simple spectral model and calculate all fluxes assuming that the source spectrum is similar to that of the Crab nebula (for which we used the absorbed power law model from \citealt{Madsen2017}). The source flux and its uncertainty intervals are estimated using Bayesian inference for the observed count rate per pixel given the measured background count rate and best estimates for the instrument PSF and vignetting.

\begin{table}[]
\footnotesize
    \caption{X-ray follow-up of \srgt. We tabulate the instrument, average date of the observation (we binned multiple visits of \nicer\ obtained within $\approx 4$ days), the estimated count rate and the total exposure time (ET).}
    \centering
    \begin{tabular}{ccccc}
    \hline
    \hline
Instrument & Observation ID & MJD & Count Rate & ET \\
          & & & ct\,s$^{-1}$ & ks \\
         \hline
\nicer\ & 4202090101-2 & 59309.67 & $2.18 \pm 0.01$ & 11.5\\
\swift\ XRT & 14257001 & 59312.87 & $0.08 \pm 0.01$  & 0.9 \\
\swift\ XRT & 14257002 & 59317.50 & $0.09 \pm 0.01$  & 0.7 \\
\nustar\ & 90701314002 & 59324.47 & $1.33 \pm 0.01$  & 36.3 \\
\swift\ XRT & 14257004 & 59329.07 & $0.32 \pm 0.03$  & 0.7\\
\swift\ XRT & 14257005 & 59333.70 & $0.18 \pm 0.02$  & 0.9\\
\swift\ XRT & 14257006 & 59338.33 & $0.09 \pm 0.01$  & 0.8 \\
\swift\ XRT & 14257008 & 59352.22 & $0.22 \pm 0.07$  & 0.2\\
\swift\ XRT & 14257009 & 59356.85 & $0.45 \pm 0.04$  & 0.9 \\
\nicer\ & 4202090103-6 & 59358.95 & $4.06 \pm 0.03$ & 8.2\\
\swift\ XRT & 14257011 & 59366.11 & $0.27 \pm 0.03$  & 0.6\\
\nicer\ & 4202090107-8 & 59374.19 & $10.14 \pm 0.04$ & 7.5\\
\swift\ XRT & 14257012 & 59393.88 & $0.61 \pm 0.06$  & 0.4 \\
\swift\ XRT & 14257014 & 59449.44 & $0.03 \pm 0.01$  & 0.9\\
\swift\ XRT & 14257015 & 59461.01 & $0.07 \pm 0.01$  & 0.8\\
\swift\ XRT & 14257016 & 59474.90 & $0.03 \pm 0.01$  & 1.0\\
\swift\ XRT & 14257017 & 59488.79 & $0.13 \pm 0.03$  & 0.2\\
\swift\ XRT & 14257018 & 59493.42 & $0.02 \pm 0.01$  & 0.7\\
\hline
    \end{tabular}

    \label{tab:xrt_lc}
\end{table}

\subsection{\swift\ Follow-up}
\label{sec:xrt_follow-up}

Following the initial detection of \srgt\ with \swift\ \citep{Heinke2021}, we requested Target of Opportunity (ToO) monitoring observations (Target ID 14257; PI: De) using the X-ray Telescope (XRT; \citealt{Burrows2005}) on board {\it Swift} \citep{Gehrels2004}. We generated the X-ray light curve of the source using the automated online tool\footnote{\url{swift.ac.uk/user_objects/}} \citep{Evans2009}. The event data were binned over $\approx 2$\,days to track the long-term X-ray evolution of the source, and the light curve is shown in Figure \ref{fig:srg_lc}. The observation log and derived count rates are given in Table \ref{tab:xrt_lc}. In addition, we examined images from the Ultraviolet Optical Telescope (UVOT; \citealt{Roming2005}) obtained during the XRT observations. The source is undetected in all of the UVOT images, as expected given its highly reddened nature. Upper limits are listed in Table \ref{tab:oirlc}.

\subsection{\nicer\ Follow-up}

\srgt\ was observed by the X-ray Timing Instrument (XTI) on board the Neutron Star Interior Composition Explorer (\nicer; \citealt{Gendreau2016}) shortly after discovery \citep{Bult2021}. Following the brightening of the X-ray transient detected in \swift, we requested two additional epochs of ToO follow-up around the peak of the X-ray flare via DDT requests. Table \ref{tab:xrt_lc} provides an overview of the \nicer\ observations. We started our data reduction with the unfiltered events lists ({\it ufa} files), which we processed using the {\tt nicerl2} task as per the guidelines given in the \nicer\ data analysis threads\footnote{\url{https://heasarc.gsfc.nasa.gov/docs/nicer/analysis_threads/}}. We used HEASoft version 6.29c with {\it NICER} calibration data (xti20210707). For {\tt nicerl2}, we used default values for all the parameters except for {\it underonly\_range}, {\it overonly\_range}, and {\it overonly\_expr} which were set to {\it `-'}, {\it `-'}, and {\it NONE}, respectively. This retains all Good Time Intervals (GTIs) that could have high optical light leak and/or particle background but are later screened based on the criteria that $hbgcut < 0.05$ and $s0cut < 2.0$ as recommended by \cite{Remillard2021}. Here, {\it hbgcut} and {\it s0cut} represent the {\emph net} count rates in the $13-15$ keV and $0.0-0.2$ keV bands, respectively, after subtracting the background contribution. 

Allowing data with all possible values of undershoots and overshoots initially and screening later based on {\it hbgcut} and {\it s0cut} values would ensure that we are not inadvertently excluding any good exposures. This procedure yielded a total exposure of roughly 26 ks. {\it NICER} consists of 52 co-aligned detectors known as Focal Plane Modules (FPMs). During a given GTI, some detectors can be `hot/bad' in the sense that they behave anomalously compared to the rest of the FPMs. This can happen temporarily due to optical light leakage into these detectors or due to other factors.  A given detector is marked as hot/bad for a given GTI if: 1) its observed low energy noise peak, i.e., $0.0-0.2$\,keV count rate, is more than four standard deviations above the median $0.0-0.2$\,keV value or 2) the observed in-band $0.4-1.0$\,keV count rate is more than two standard deviations above the median $0.4-1.0$\,keV of all detectors. We then generate combined responses for the GTI excluding these hot/bad detectors. We use the \texttt{ni3C50} background model \citep{Remillard2021} to estimate the background. The final background subtracted light curve normalized to 50 FPMs is shown in Figure \ref{fig:srg_lc}.

\subsection{\nustar\ Follow-up}

We obtained one epoch of DDT ToO follow-up (PI: K. De; Sequence ID 90701314002) of \srgt\ using the \nustar\ hard X-ray telescope \citep{Harrison2013}. Table \ref{tab:xrt_lc} provides a summary of the \nustar\ observation. The data from the two focal plane modules were processed using the \nustar\ Data Analysis Software (\texttt{nustardas}) V2.1.1 and the 20211202 CALDB. The \texttt{nupipeline} command was used to produce cleaned event files using the default parameters. Barycentric correction was performed on the event files using the \texttt{barycorr} task and the  clock offset correction file version V128 \citep{Bachetti2021}. Light curves and spectra for the source were generated by extracting events within a radius of 60\arcsec\ centered at the position of the source using \texttt{nuproducts}. Background events were selected from a region of 100\arcsec\ located away from the source in the same part of the detector. Light curves were generated with the \texttt{Stingray} package V0.3 \citep{Stingray} using the barycentered event lists.

\section{The X-ray Characteristics}
\label{sec:xrayprop}

\subsection{Broadband spectral analysis}
\label{sec:xray_spectra}

\begin{table}[!ht]

    \scriptsize
    \centering
    
  Model 2: \texttt{edge * Tbabs * (pcfabs * (highecut * powerlaw + gauss) + apec)}\\
  $\chi^2$/d.o.f. = $257.9/226$, P$_{\rm null} =0.06$\\
    \begin{tabular}{cccc}
    
    \hline
         Component & Parameter & Value & Unit \\
         \hline
         \texttt{const} & $C_{Ni}$ & $0.88^{+0.02}_{-0.02}$ & - \\
         \texttt{const} & $C_{NuB}$ & $1.04^{+0.02}_{-0.02}$ & - \\
         \texttt{edge} & $E$ & $1.844$ (fixed) & keV\\
         \texttt{edge} & $\tau$ & $0.30^{+0.06}_{-0.08}$ & - \\
         \texttt{Tbabs} & $N_{H}$ & $0.6^{+0.1}_{-0.1}$ & $10^{22}$\,cm$^{-2}$\\
         \texttt{pcfabs} & $N_{H}$ & $5.8^{+0.4}_{-0.4}$ & $10^{22}$\,cm$^{-2}$\\
         \texttt{pcfabs} & $CF$ & $0.95^{+0.01}_{-0.01}$ & -\\
         \texttt{highecut} & $E_{cut}$ & $4.8^{+0.4}_{-0.4}$ & keV \\
         \texttt{highecut} & $E_{fold}$ & $12^{+1}_{-1}$ & keV\\
         \texttt{powerlaw} & $\Gamma$ & $1.5^{+0.1}_{-0.1}$ & - \\
         \texttt{gauss} & LineE & $6.44^{+0.04}_{-0.04}$ & keV \\
         \texttt{gauss} & Sigma & $9^{+6}_{-5} \times 10^{-2}$ & keV\\
         \texttt{apec} & kT & $0.62^{+0.04}_{-0.04}$ & keV\\
         \texttt{apec} & norm & $4^{+1}_{-1} \times 10^{-4}$ & cm$^{-5}$\\
         \hline
    \end{tabular}
    
    \vspace{0.3cm}
    
 Model 3: \texttt{edge * Tbabs * (pcfabs * (compTT + gauss) + apec)}\\
 $\chi^2$/d.o.f. = $250.4/223$, P$_{\rm null}=0.10$\\
    \begin{tabular}{cccc}
    
    \hline
         Component & Parameter & Value & Unit \\
         \hline
         \texttt{const} & $C_{Ni}$ & $0.89^{+0.02}_{-0.02}$ & - \\
         \texttt{const} & $C_{NuB}$ & $1.04^{+0.02}_{-0.02}$ & - \\
         \texttt{edge} & $E$ & $1.844$ (fixed) & keV\\
         \texttt{edge} & $\tau$ & $0.17^{+0.23}_{-0.07}$ & - \\
         \texttt{Tbabs} & $N_{H}$ & $0.8^{+0.1}_{-0.1}$ & $10^{22}$\,cm$^{-2}$\\
         \texttt{pcfabs} & $N_{H}$ & $2^{+1}_{-1}$ & $10^{22}$\,cm$^{-2}$\\
         \texttt{pcfabs} & $CF$ & $0.8^{+0.2}_{-0.1}$ & -\\
         \texttt{compTT} & $T0$ & $1.03^{+0.03}_{-0.04}$ & keV\\
         \texttt{compTT} & $kT$ & $8^{+3}_{-1}$ & keV \\
         \texttt{compTT} & $\tau$ & $6.1^{+0.9}_{-0.8}$ & - \\
         \texttt{gauss} & LineE & $6.44^{+0.04}_{-0.04}$ & keV \\
         \texttt{gauss} & Sigma & $7^{+5}_{-5} \times 10^{-2}$ & keV\\
         \texttt{apec} & kT & $0.62^{+0.03}_{-0.03}$ & keV \\
         \texttt{apec} & norm & $8^{+3}_{-2} \times 10^{-4}$ & cm$^{-5}$\\
         \hline
    \end{tabular}
    \caption{Comparison of the best-fit parameters used to model the broadband X-ray spectrum of \srgt. For each model, the \nustar\ FPMA constant was set to 1 while the scaling constants for the \nicer\ and \nustar\ FPMB spectra were allowed to vary. We show the best-fit $\chi^2$ values and null hypothesis probability for each model above the parameter table. $CF$ denotes the covering fraction for the \texttt{pcfabs} model. The \texttt{apec} model norm is related to the emission measure ($EM$) of the ionized region as $\texttt{norm} = \frac{10^{-14}}{4 \pi D^2} EM$, where $D$ is the distance to the source.}
    \label{tab:xmodels}
\end{table}

We begin the discussion on the X-ray properties of \srgt\ with the broadband spectral analysis. We use the \nicer\ and \nustar\ observations near the first detection of the outburst in April 2021 (see Table \ref{tab:xrt_lc}) to perform a joint fit. Although not strictly simultaneous, there is little spectral variability between the \nicer\ and \nustar\ observations and we fit them together to well constrain the broadband continuum and absorption column density. The fitting was carried out using the \texttt{xspec} package \citep{Arnaud1996} version 12.12.0. For \nustar, the individual FPMA and FPMB spectra produced by \texttt{nuproducts} were grouped with \texttt{ftgrouppha} to have at least 20 counts per bin in order to fit them with $\chi^2$ statistics. We included  data in the $3.0 - 30.0$\,keV energy range since the spectrum/observation becomes background dominated at higher energies. The \nicer\ data were similarly binned to include at least 20 counts per bin. We include \nicer\ data in the 0.7-8.0\,keV range since the source is heavily absorbed at lower energies and comparable to the background at higher energies. Uncertainties of model parameters are presented at the 68\% confidence level. The derived parameters for the models are discussed below and given in Table \ref{tab:xmodels}.

\begin{figure*}[!ht]
    \centering
    \includegraphics[width=\textwidth]{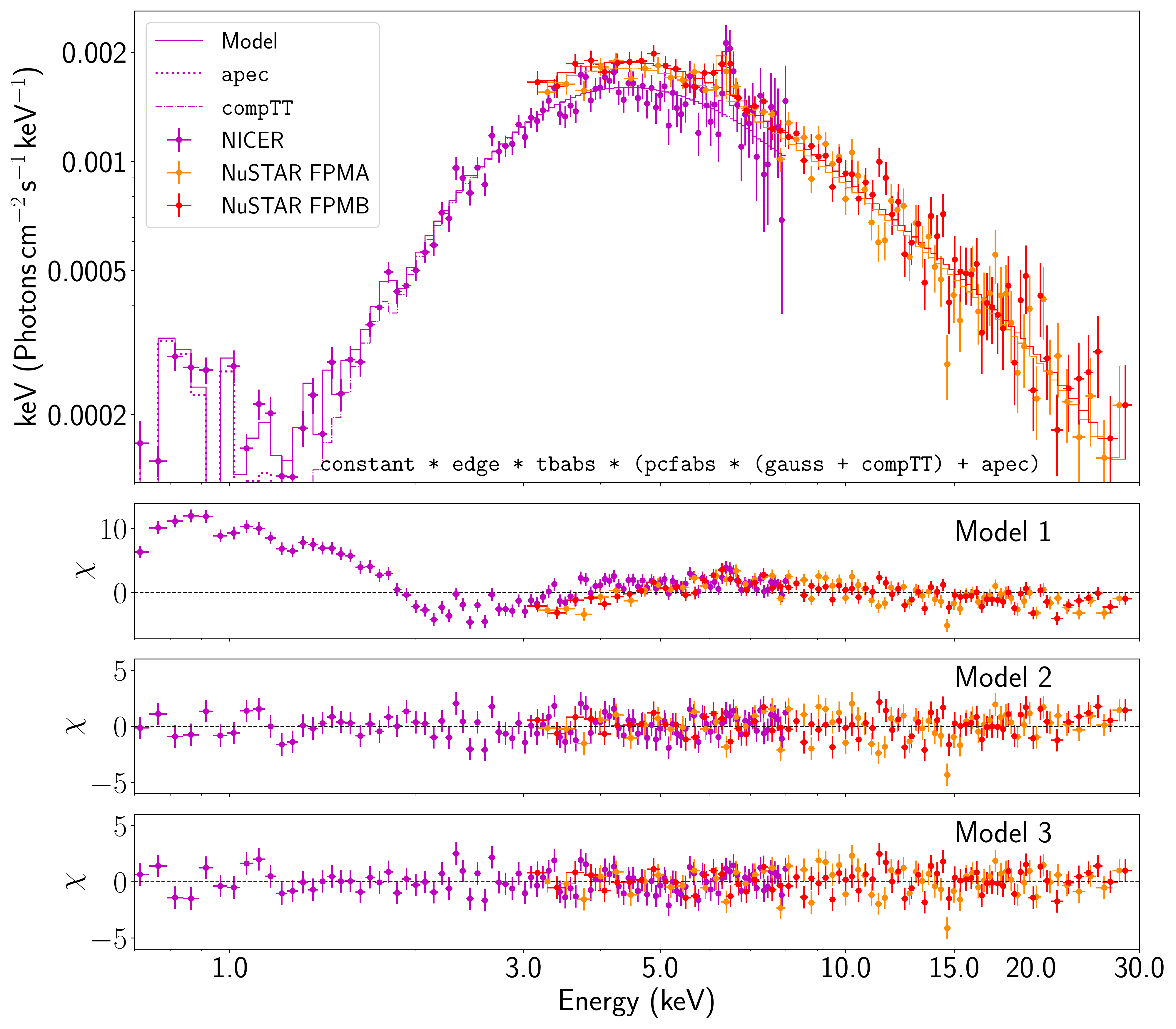}
    \caption{Broadband X-ray spectral modeling of \srgt. The top panel shows the data along with the preferred physical model consisting of a Comptonized plasma with seed photons, a lower energy blackbody component and a 6.4 keV Fe line. The \texttt{bbody} and \texttt{compTT} model contributions to the final model are shown separately. The lower panels show residuals with respect to the different models attempted, as listed in Table \ref{tab:xmodels}. Model 1 is a simple absorbed power law model which does not provide a good description of the data (see text).}
    \label{fig:xspec}
\end{figure*}

Preliminary analysis of the X-ray spectra suggested a heavily absorbed power law spectrum for \srgt\ \citep{Mereminskiy2021, Bult2021, Heinke2021}. \citet{Bult2021} also noted a low energy excess in the NICER spectra which was attributed to a possible unrelated source in the field of view. However, the \swift\ XRT images do not show any other source within the NICER field. We thus first attempted to fit the $0.7-30.0$ keV spectrum using a simple absorbed power-law model (\texttt{Tbabs * powerlaw}) using the \texttt{wilms} interstellar abundances \citep{Wilms2000}. We allowed for a varying normalization between \nicer\ and \nustar, and the standard cross-calibration constant between the two \nustar\ modules. The resulting model is \texttt{constant * Tbabs * powerlaw} (Model 1). We performed spectral fitting with $\chi^2$ minimization, and the resulting residuals are shown in Figure \ref{fig:xspec}. The best-fit model suggests a high column density of $N_{\rm H} \approx 7.2 \times 10^{22}$\,cm$^{-2}$ consistent with preliminary reports, and a photon index of $\Gamma \approx 2.2$. However, the power-law fit is overall poor with $\chi^2$/d.o.f $\approx $ 2097.84/239, and leaves clear positive residuals below $\approx 2$\,keV and between $\approx 5-10$\,keV, and negative residuals above $\approx 15$\,keV. The residuals also reveal a feature at 6.4\,keV. We do not discuss this model further.

We then attempted a more complex phenomenological model consisting of a cuttoff power-law implemented using \texttt{highecut * powerlaw} for the high energy emission component, where the high energy roll over ($E_{cut}$) for the power law takes effect at a variable cut-off ($E_{fold}$) energy (Model 2). We used an \texttt{apec} \citep{Smith2001} component to model the low energy excess\footnote{Due to the very high absorption column, alternative attempts at fitting the low energy component with blackbody (\texttt{bbody}) or disk (\texttt{diskbb}) emission produced fits that were poorer or produced unreasonably large luminosity/radii for optically thick components.}. Corroborating evidence for the presence of an optically thin plasma component is clearly detected in subsequent \nicer\ spectra, and is discussed in Section \ref{sec:xspecevol}. We fixed the \texttt{apec} abundance to $1.0$, thus assuming a \citet{Wilms2000} composition.  We found improved fits by adding a partial absorber (\texttt{pcfabs}) to the high energy component in addition to \texttt{Tbabs}\footnote{Placing the \texttt{apec} and high energy component behind the same absorbing column produces unphysically high luminosity for the \texttt{apec} component. We therefore suggest that the high energy emission component exhibits a higher absorption column.}.

A clear 6.4\,keV excess is also seen in the residuals from the simple power-law fit, which we model using a \texttt{gauss} component in the fit. Due to limitations of the F-test for line emission features \citep{Protassov2002}, we simulate its significance using the \texttt{simftest} command in \texttt{xspec}, which suggests a null-hypothesis probability of $<0.01$. We further note a systematic feature at $\approx 1.8-1.9$\,keV, which is clearly detected as an absorption edge in subsequent \nicer\ spectra at higher fluxes (see Section \ref{sec:xspecevol}). We identify this feature as the Si K-edge at 1.844\,keV, which is detected in many highly absorbed X-ray binaries \citep{Schulz2016}. We include this feature in the fit using the \texttt{edge} component in \texttt{xspec}, keeping the energy fixed and the optical depth as a variable parameter. The addition of the edge indeed results in an improved fit; simulations for the F-test suggest a null-hypothesis probability of $<0.03$. The resulting model (Model 2) residuals are shown in Figure \ref{fig:xspec}, and provides a better fit than the power law model with $\chi^2$/d.o.f.$= 257.9/225$ (Table \ref{tab:xmodels}).

Since these models are empirical in nature, we tried a physically motivated interpretation by fitting a Comptonization model, \texttt{compTT} \citep{Titarchuk1995}.  The best-fit model (Model 3) indicates a seed photon temperature of $\approx 1.0$\,keV and a plasma temperature of $\approx 7.5$\,keV for the \texttt{comptt} component, while the low energy \texttt{apec} component has a temperature of $\approx 0.6$\,keV. The model provided an acceptable fit ($\chi^2$/d.o.f.$= 250.43/223$).  The derived parameters for the Comptonized plasma are similar to several known SyXRBs \citep{Masetti2002, Masetti2007, Enoto2014, Bozzo2018}. We thus adopt this as the working model for \srgt. We compute the integrated absorbed and unabsorbed X-ray fluxes of the source using the \texttt{cflux} component in \texttt{xspec} added to the best-fit spectral model. Placing the \texttt{cflux} component before the foreground \texttt{Tbabs} absorber, we obtain absorbed fluxes of $1.33^{+0.02}_{-0.01}\times 10^{-11}$\,erg\,cm$^{-2}$\,s$^{-1}$ and $2.16^{+0.03}_{-0.03}\times 10^{-11}$\,erg\,cm$^{-2}$\,s$^{-1}$ in the $0.8-8.0$ and $3.0-30.0$\,keV range respectively. Placing the \texttt{cflux} component after the \texttt{Tbabs} absorber, we derive unabsorbed X-ray fluxes of $1.67^{+0.03}_{-0.03}\times 10^{-11}$\,erg\,cm$^{-2}$\,s$^{-1}$ and $2.48^{+0.04}_{-0.04}\times 10^{-11}$\,erg\,cm$^{-2}$\,s$^{-1}$ in the $0.8-8.0$ and $3.0-30.0$\,keV range respectively.

\subsection{Spectral evolution}
\label{sec:xspecevol}
\begin{figure}
    \centering
    \includegraphics[width=\columnwidth]{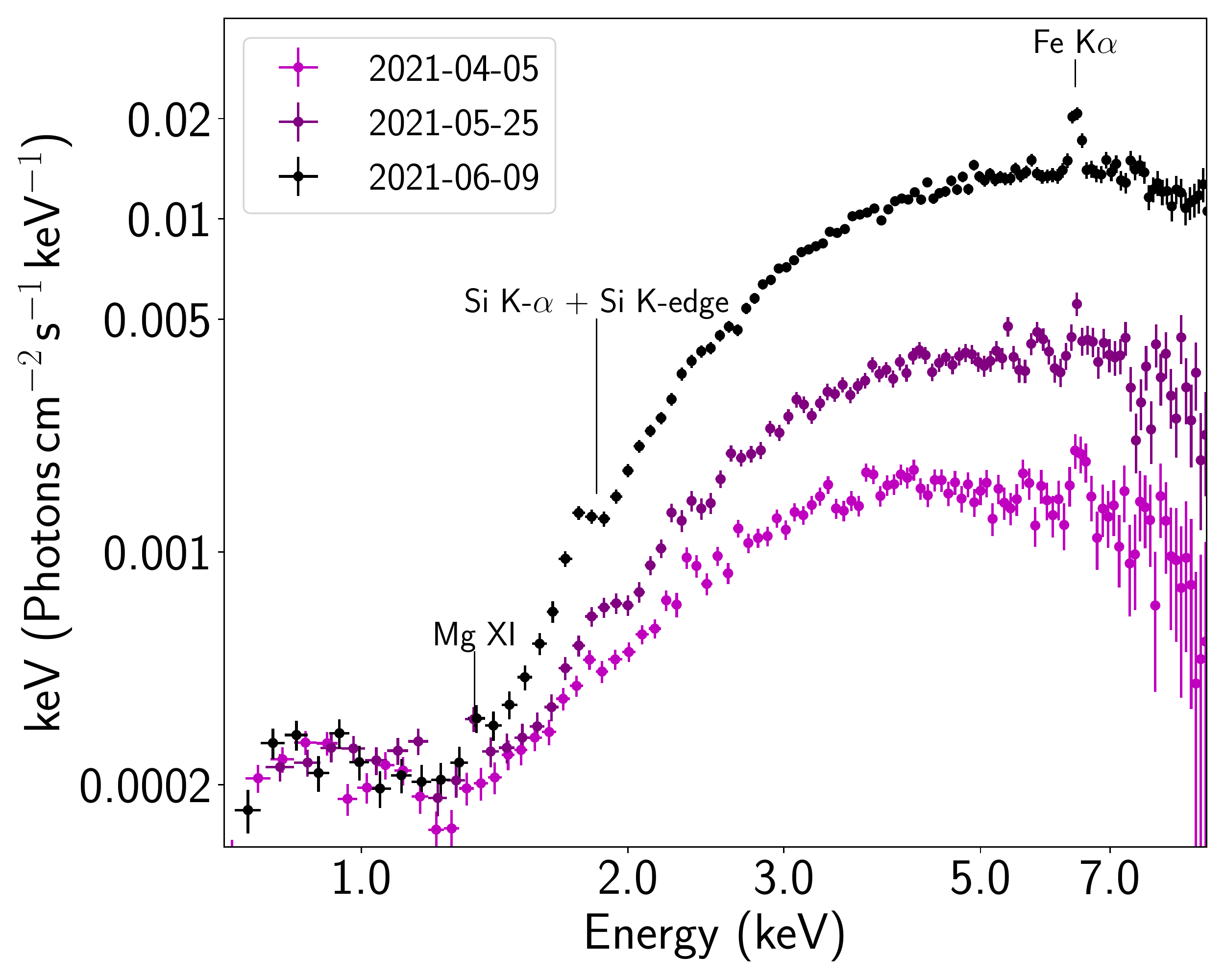}
    \caption{Spectral evolution of \srgt\ during the 2021 outburst observed by \nicer. We show the unfolded $F_\nu$ spectra assuming a flat power law model with $\Gamma = 0$ to avoid artificially introducing sharp features from the model assumption. Prominent atomic features are marked.}
    \label{fig:specevol}
\end{figure}

\begin{table}[]

\footnotesize
    \centering
    \nicer\ Epoch 2 (2021-05-25):\\
     $\chi^2$/d.o.f. = $126.67/95$, P$_{\rm null}=0.02$\\
      $F_X = 3.23^{+0.04}_{-0.03} \times 10^{-11}$\,erg\,cm$^{-2}$\,s$^{-1}$\\
    \begin{tabular}{cccc}
    \hline
         Component & Parameter & Value & Unit \\
         \hline
         \texttt{edge} & $\tau$ & $0.13^{+0.09}_{-0.09}$ & - \\
         \texttt{Tbabs} & $N_{H}$ & $1.2^{+0.1}_{-0.1}$ & $10^{22}$\,cm$^{-2}$\\
         \texttt{pcfabs} & $N_{H}$ & $5^{+1}_{-1}$ & $10^{22}$\,cm$^{-2}$\\
         \texttt{pcfabs} & $CF$ & $0.92^{+0.08}_{-0.05}$ & -\\
         \texttt{compTT} & $T0$ & $0.6^{+0.4}_{-0.4}$ & keV\\
         \texttt{compTT} & $kT$ & $2.4^{+0.3}_{-0.2}$ & keV \\
         \texttt{compTT} & $\tau$ & $21^{+3}_{-3}$ & - \\
         \texttt{gauss} & LineE & $6.42^{+0.02}_{-0.02}$ & keV \\
         \texttt{gauss} & Sigma & $1.1^{+5.1}_{-0.8} \times 10^{-2}$ & keV\\
         \texttt{apec} & kT & $0.58^{+0.04}_{-0.05}$ & keV \\
         \texttt{apec} & norm & $18^{+5}_{-5} \times 10^{-4}$ & cm$^{-5}$\\
        \hline
    \end{tabular}
    
    \vspace{0.3cm}
    
    \nicer\ Epoch 3 (2021-06-29):\\
     $\chi^2$/d.o.f. = $115.68/102$, P$_{\rm null}=0.17$\\
      $F_X = 10.96^{+0.07}_{-0.04} \times 10^{-11}$\,erg\,cm$^{-2}$\,s$^{-1}$\\
    \begin{tabular}{cccc}
    \hline
         Component & Parameter & Value & Unit \\
         \hline
         \texttt{edge} & $\tau$ & $0.29^{+0.03}_{-0.07}$ & - \\
         \texttt{Tbabs} & $N_{H}$ & $1.05^{+0.07}_{-0.08}$ & $10^{22}$\,cm$^{-2}$\\
         \texttt{pcfabs} & $N_{H}$ & $4.21^{+0.05}_{-0.51}$ & $10^{22}$\,cm$^{-2}$\\
         \texttt{pcfabs} & $CF$ & $0.99^{+0.01}_{-0.07}$ & -\\
         \texttt{compTT} & $T0$ & $0.61^{+0.01}_{-0.14}$ & keV\\
         \texttt{compTT} & $kT$ & $2.45^{+0.02}_{-0.12}$ & keV \\
         \texttt{compTT} & $\tau$ & $26^{+1}_{-5}$ & - \\
         \texttt{gauss} & LineE & $6.39^{+0.01}_{-0.01}$ & keV \\
         \texttt{gauss} & Sigma* & $4^{+1}_{-2} \times 10^{-2}$ & keV\\
         \texttt{apec} & kT & $0.59^{+0.05}_{-0.02}$ & keV \\
         \texttt{apec} & norm & $10.3^{+2.6}_{-0.5} \times 10^{-4}$ & cm$^{-5}$\\
         \hline
    \end{tabular}   
    \caption{Evolution of the best-fit model parameters for \srgt\ as observed with \nicer. The parameters correspond to the adopted Model 3 in Table \ref{tab:xmodels}. For each model, we show the best fit parameters, their uncertainties, the resulting $\chi^2$ as well as estimated absorbed $0.8-8.0$\,keV fluxes (as $F_X$).}
    \label{tab:nimodel_evol}

\end{table}

We attempted to fit the two subsequent epochs of \nicer\ spectra using the same Comptonization model discussed in the joint analysis and were able to get acceptable fits. We list the derived parameters in Table \ref{tab:nimodel_evol}, noting that the high energy \texttt{comptt} component is relatively poorly constrained due to the absence of higher energy coverage. In Figure \ref{fig:specevol}, we show the \nicer\ spectra in the $0.9 - 9.0$\,keV range for the three epochs. The low energy \texttt{apec} component is present in all the spectra, while we see a clear hardening of spectrum with increasing X-ray flux. Similar hardening of the X-ray spectrum with increasing flux have been previously observed in other SyXRBs \citep{Masetti2002, Enoto2014}.

The subsequent \nicer\ observations also show evidence of unresolved emission and absorption features that are well detected in the higher signal-to-noise ratio spectra. In the data taken around UT 2021-05-25, we find an unresolved emission line around 1.34\,keV, which coincides with a known feature of Mg\,XI. The feature is strongest at this epoch (Figure \ref{fig:specevol}), but is also visible in the other two epochs at lower significance. The variability argues against a possible origin in background Galactic ridge X-ray emission \citep{Revnivtsev2006}. This emission line suggests the presence of an optically thin plasma emission component, and is well fit by the adopted \texttt{apec} component. The \nicer\ spectrum near the peak of the 2021 outburst (around UT 2021-06-09) shows a strong absorption feature near the Si K-edge, while a lower significance feature is detected in the earlier epochs. The derived optical depth of the Si edge is variable across the epochs, although we cannot rule out unresolved contamination with the nearby Si K$\alpha$ emission line in these low resolution spectra.

There is clear variability in the narrow Fe K$\alpha$ line. The line is detected in all epochs as confirmed with the F-test, which suggests a null-hypothesis probability of $\approx 3 \times 10^{-3}$ and $\approx 10^{-11}$ in the second and third epochs respectively. The line center is consistent with rest wavelength and is unresolved (\texttt{Sigma}$< 140$\,eV; the spectral resolution of \nicer). This suggests that it does not arise from reflection from the inner part of the disk \citep{Garcia2011}; it may come from reflection off colder gas at larger distances, or from the extended thermally-emitting plasma. Based on the best-fit parameters, we find the broadband hardening to be associated with increasing optical depth for the Comptonizing plasma towards the peak of the outburst. We also note a variable covering fraction for the partial absorber, while the \texttt{Tbabs} absorber remains roughly constant.

\subsection{Temporal variability}

\begin{figure}
    \centering
    \includegraphics[width=\columnwidth]{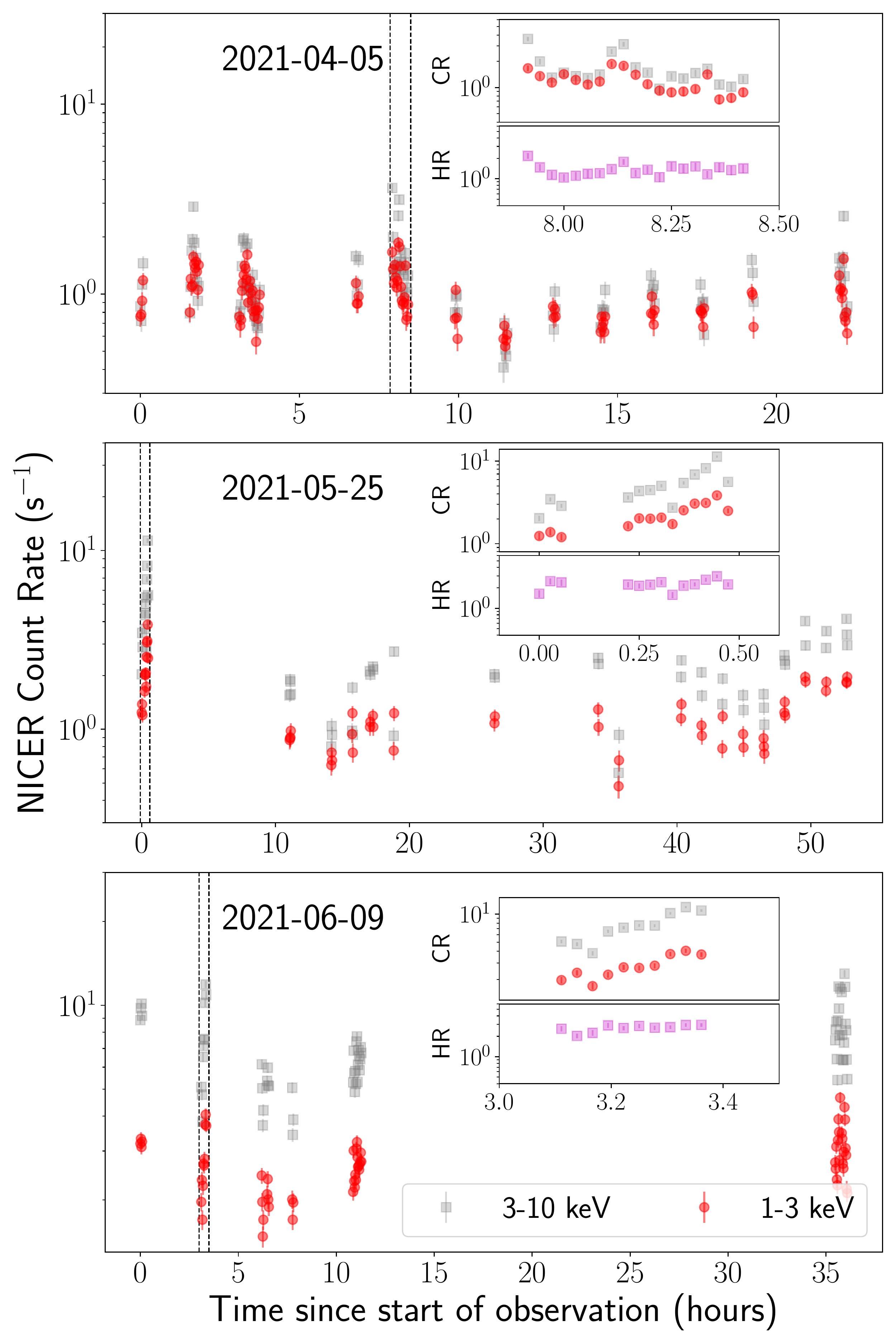}
    \caption{Short term variability in \srgt\ observed with \nicer. In each panel (showing the mean date of observation), we plot the light curve binned to 100\,s time resolution over multiple visits of \nicer. The insets show zoom-ins of the light curves around time ranges (marked with vertical dashed lines) with high X-ray activity. In each of the insets, we also show the variability of the hardness ratio (HR; in magenta) below the count rate (CR) light curve.}
    \label{fig:nicerlc}
\end{figure}

\begin{figure}
    \centering
    \includegraphics[width=\columnwidth]{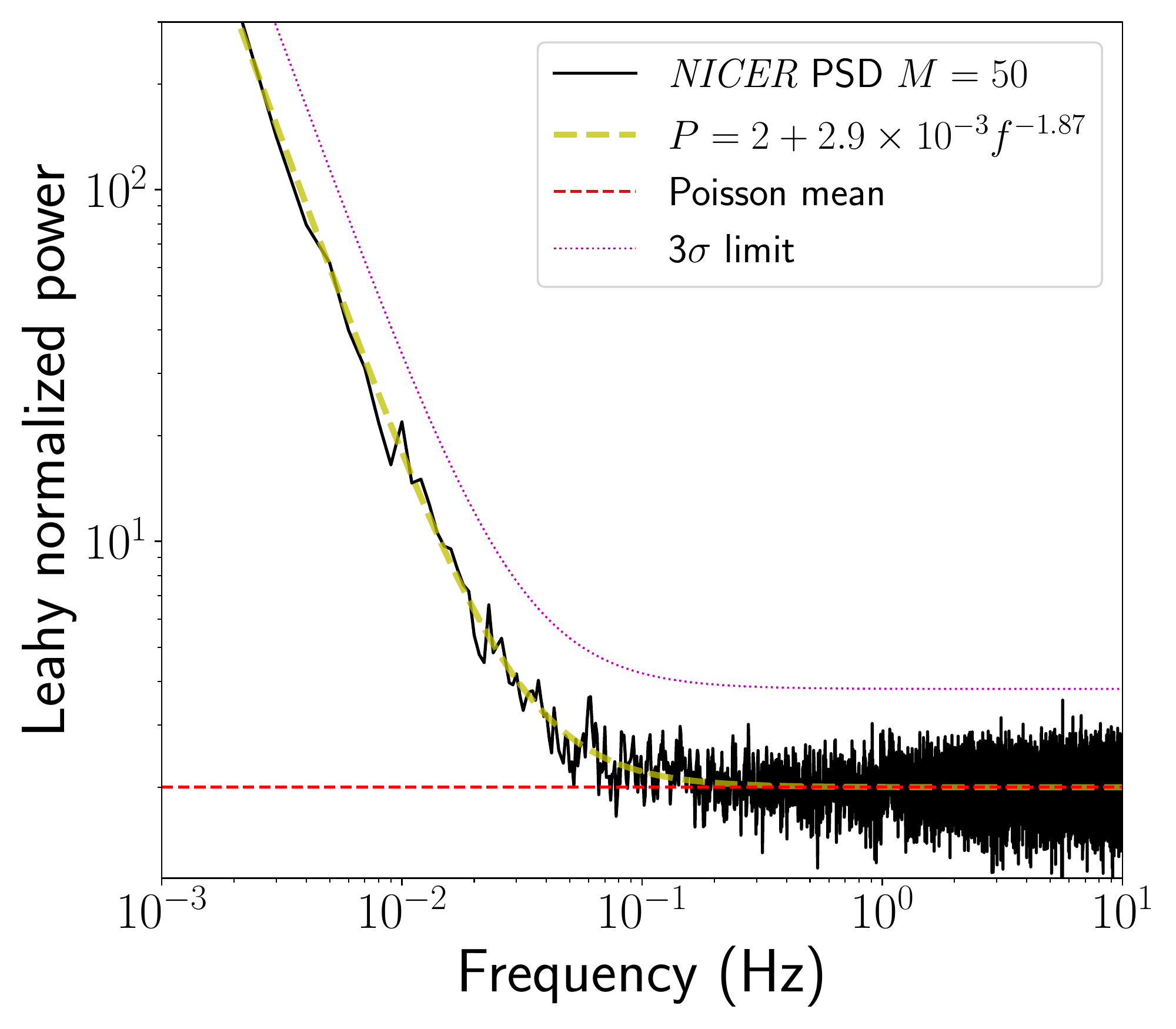}
    \caption{Normalized power spectrum of the combined \nicer\ dataset, obtained by averaging $M = 50$ segments of 1000\,s each. The Poisson mean curve shows the expected power value for pure white noise. We also show the best-fit power law description of the red noise component, together with the $3\sigma$ upper limit curve for the presence of a pulsation superimposed on the red noise. }
    \label{fig:nicer_psd}
\end{figure}

\srgt\ exhibits dramatic X-ray variability on timescales of minutes to weeks. As in Figure \ref{fig:srg_lc} and Section \ref{sec:xspecevol}, the X-ray transient was observed to brighten and harden for $\approx 100$\,days after the initial discovery, followed by a drop to its near-quiescent flux level over the next $\approx 100$\,days. The individual \nicer\ observations obtained during the brightening provide strong constraints on the short-term variability. Figure \ref{fig:nicerlc} shows the X-ray light curve of the source in the soft ($1-3$\,keV) and hard ($3-10$\,keV) bands during the three epochs of \nicer\ observations. All epochs are characterized by brief periods of rapid X-ray activity where the source flux varies by $\gtrsim 5\times$ over timescales of $\sim$minutes. The short timescale variability is nearly identical in the hard and soft bands, as indicated by the relatively constant hardness ratio as the count rate changes by factor of $\approx 10\times$. Similar `colorless' variability is known in several SyXRBs \citep{Masetti2002, Enoto2014}. We also searched for the presence of X-ray bursts by binning the \nicer\ light curve at $1$\,s time resolution but did not find any bursts.

To search for the presence of pulsed emission, we used the complete \nicer\ event list (corrected to barycentric time relative to the JPLEPH.DE200 ephemeris as per the default in the FK5 reference frame) and created a mean power spectral density (PSD) from $10^{-3}$\,Hz to $10$\,Hz. We averaged power spectra of segments of 1000\,s each, with a time resolution of $0.1$\,s. Figure \ref{fig:nicer_psd} shows that the source exhibits a strong red noise dominated PSD. We searched for the presence of a pulsation by fitting the red noise component with a power law and searching for excess power after normalizing the PSD by the red noise component \citep{vanderKlis1989, Israel1996}. The magenta dotted line in Figure \ref{fig:nicer_psd} shows the $3\sigma$ threshold for the detection of a signal accounting for the expected noise distribution in the averaged PSD. No pulsations are detected at the $3\sigma$ level, corresponding to an upper limit on the fractional pulsation amplitude of $\sim 1$\% at $\sim 10$\,Hz and $\sim 20$\% at $\sim 10^{-3}$\,Hz. A similar pulsation analysis on the \nustar\ data also did not reveal pulsations at the $3\sigma$ level, although the constraints are less stringent due to the lower number of counts.

\section{The Infrared Counterpart}
\label{sec:ircounterpart}

\subsection{Localization and Association}

At the time of discovery, the best-fit position of \srgt\ was reported from the eROSITA instrument with an error radius of $\approx 9$\,\arcsec, which included the position of the variable infrared counterpart IRAS\,18111-2257. We used the \swift\ XRT data to derive a higher precision position of the X-ray source. Using all the XRT event data obtained post-discovery, we used the online XRT products tool\footnote{\url{https://www.swift.ac.uk/user_objects/}} tool \citep{Evans2009} to derive an enhanced position using stars in the UVOT field of view \citep{Goad2007}. The best-fit position is $\alpha=$ 18:14:14.74, $\delta =$ -22:56:19.1 with an error-radius of $\approx 2.4$\arcsec. Figure \ref{fig:localization} shows a NIR image of the field overlaid with the X-ray localization regions from ART-XC, eROSITA and \swift. The best-fit position is squarely consistent with the position of IRAS\,18111-2257.

IRAS\,18111-2257 is reported in the 2MASS Point Source Catalog \citep{Skrutskie2006} as 2MASS\,J18141475-2256195, with NIR magnitudes of $J = 8.32 \pm 0.07$, $H = 6.93 \pm 0.04$ and $Ks = 6.15 \pm 0.02$, and is coincident within $\approx 0.4$\arcsec\ of the position of the X-ray source. The IR source is one of the brightest and reddest sources in the nearby field, making the spatial association with \srgt\ very probable. Following \citet{Kaplan2007}, we use the 2MASS catalog to estimate a spatial density of $\approx 7\times 10^{-6}$\arcsec$^{-2}$ for sources brighter than $K_s = 6.5$ in a $10\times10$\,arcmin$^2$ region around the source. This suggests a random probability of $\approx 3 \times 10^{-6}$ for the X-ray transient to be spatially coincident within $\approx 0.4$\arcsec\ of the bright IR source. Similarly, using the PGIR archival data, we derive a chance coincidence probability of $\approx 4 \times 10^{-5}$ for the X-ray transient to be coincident within $0.4$\arcsec\ of a variable 2MASS source brighter than $J = 10.0$\,mag. Together with the temporal coincidence of the IR and X-ray brightening, we conclude that the IR source is very likely the infrared counterpart of \srgt. Hereafter, we refer to the IR counterpart of \srgt\ as IRAS\,18111-2257.

\subsection{Spectral Type}

\begin{figure*}[!htp]
    \centering
    \includegraphics[width=0.9\textwidth]{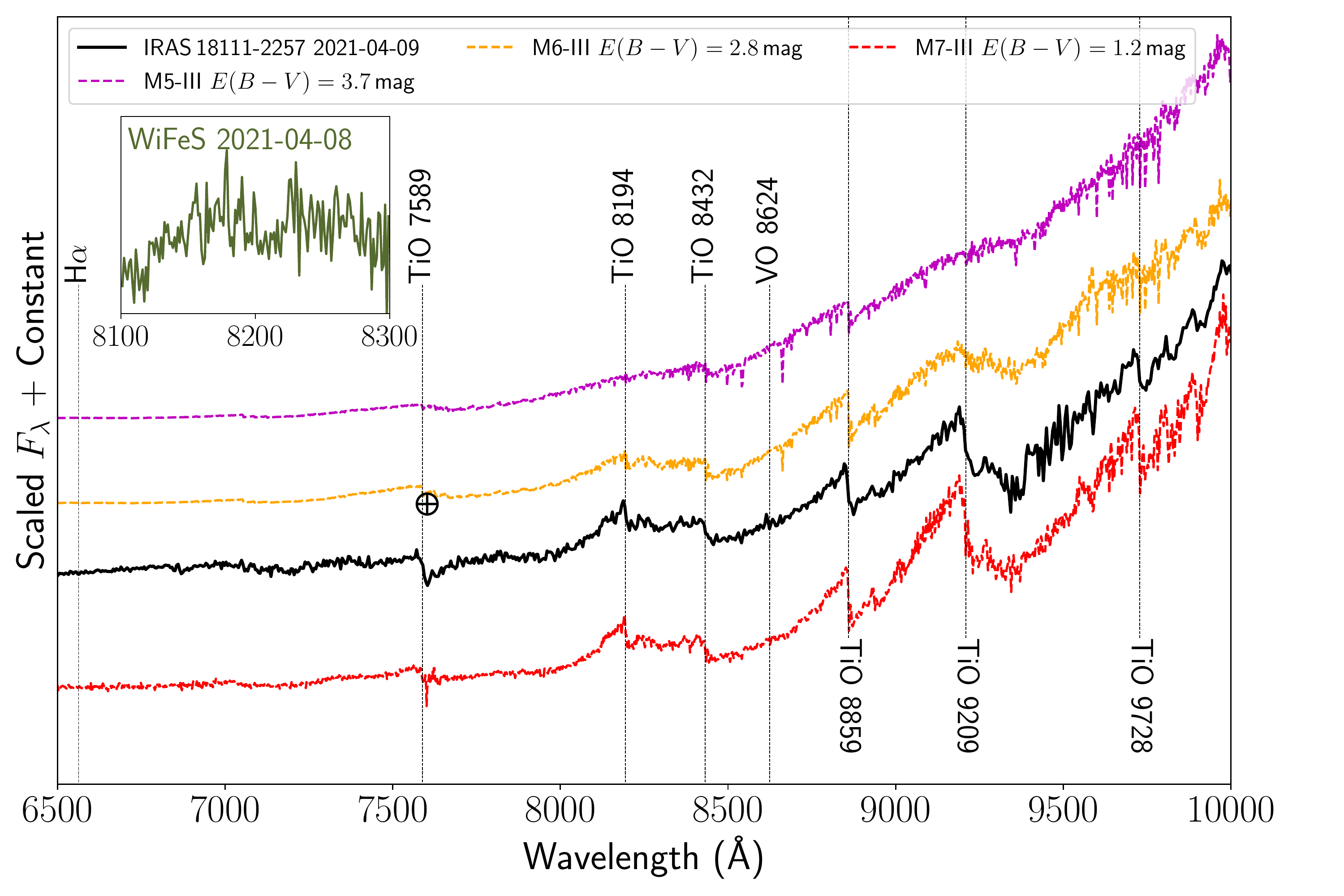}
    \caption{Comparison of the low resolution optical spectra of IRAS\,18111-2257 with spectral standards from the VLT X-Shooter Spectral Library \citep{Gonneau2020}. The standard spectra were artificially extinguished with additional reddening (using the \citealt{Cardelli1989} extinction law with $R_V = 3.1$) as indicated in the legend to match the spectral slope of IRAS\,18111-2257 for easier comparison. The inset shows a higher resolution contemporaneous optical spectrum from WiFeS, zoomed in around the region of the Na 8200\,\AA\ feature. The deep TiO and VO absorption band-heads suggest a spectral type of at least M6-M7, while the absence of strong Na 8200\,\AA\ absorption \citep{Cushing2005} constrains the luminosity class to a giant.}
    \label{fig:opt_spec}
\end{figure*}

\begin{figure}[!htp]
    \centering
    \includegraphics[width=\columnwidth]{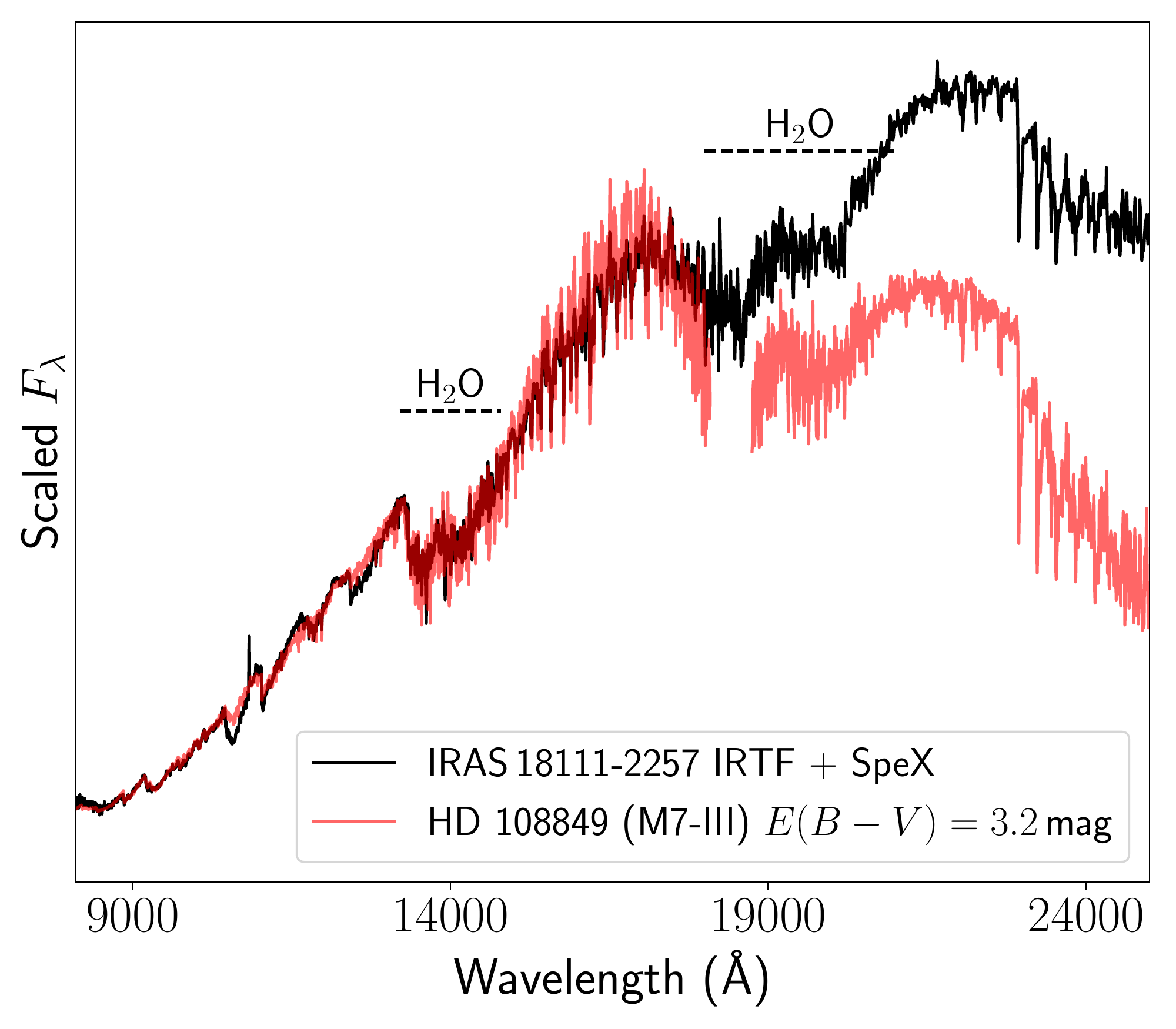}
    \caption{The NIR spectrum of IRAS\,18111-2257 compared with a star of the same spectral type (HD\,108849; as inferred from the optical spectrum) from the IRTF Library \citep{Rayner2009}. The comparison star spectrum has been reddened with extinction (indicated in legend) as in the case of the optical spectra. The reddening was applied to match the slope of the $J$ and $H$-band spectra, but consistently leaves a large excess in $K$-band (while a larger value of extinction is inconsistent with the $J$ and $H$ band continuum slope), likely suggesting the presence of warm dust. Prominent water absorption features are visible and marked.}
    \label{fig:nir_bspec}
\end{figure}

In Figure \ref{fig:opt_spec}, we show the two epochs of low resolution ($R\approx 1000$) optical spectra of IRAS\,18111-2257 obtained with DBSP. The most prominent absorption features are identified with sharp TiO and VO absorption band-heads typically seen in late M-type stars \citep{Lancon2007}. The inset in Figure \ref{fig:opt_spec} shows the higher resolution spectrum obtained with WiFeS, zoomed in around 8200\,\AA. The source does not exhibit strong Na absorption around 8200\,\AA, which is characteristic of late M-type dwarfs \citep{Cushing2005}, suggesting that the counterpart is a luminous giant or supergiant.

It is well established that the TiO band strengths correlate well with the spectral type of late-type giants, with deeper bands corresponding to later and cooler spectral types (e.g. \citealt{Ramsey1981, Kenyon1987}). To identify the spectral type, we compare our spectra to that of late type (M5-M7) optical spectra of cool giants in the recent VLT X-Shooter Spectral Library DR2 \citep{Gonneau2020} in Figure \ref{fig:opt_spec}. The steep red continuum of the source is inconsistent with the typical colors of cool giants, and hence an additional dust extinction was applied to the library spectra for easier comparison. As shown, the deep TiO absorption features suggest the spectral type to be $>$ M6-M7.

Figure \ref{fig:nir_bspec} shows the NIR spectrum of IRAS\,18111-2257 compared to a star of the same spectral type (as inferred from the optical spectrum) in the IRTF spectral library \citep{Rayner2009}. The spectrum shows clear signatures of broad water absorption features around $\approx 1.4$ and $\approx 2.0\,\mu$m, which are only seen in the very extended atmospheres of large amplitude pulsating Mira variables (amplitude $\delta V > 1.7$\,mag; \citealt{Matsuura1999,Lancon1999}). We measure the H$_2$O absorption index defined in \citet{Blum2003} to be $\approx 8$\%, suggesting that the source is a Mira AGB star since supergiants show absorption indices of $< 6$\% \citep{Messineo2021}. Compared to the reddened spectrum of the comparison star, IRAS\,18111-2257 exhibits a large excess in $K$-band, suggestive of the presence of warm circumstellar dust as seen in D-type symbiotic binaries and consistent with the steep reddened optical spectrum.

\begin{figure}[!hpt]
    \centering
    \includegraphics[width=\columnwidth]{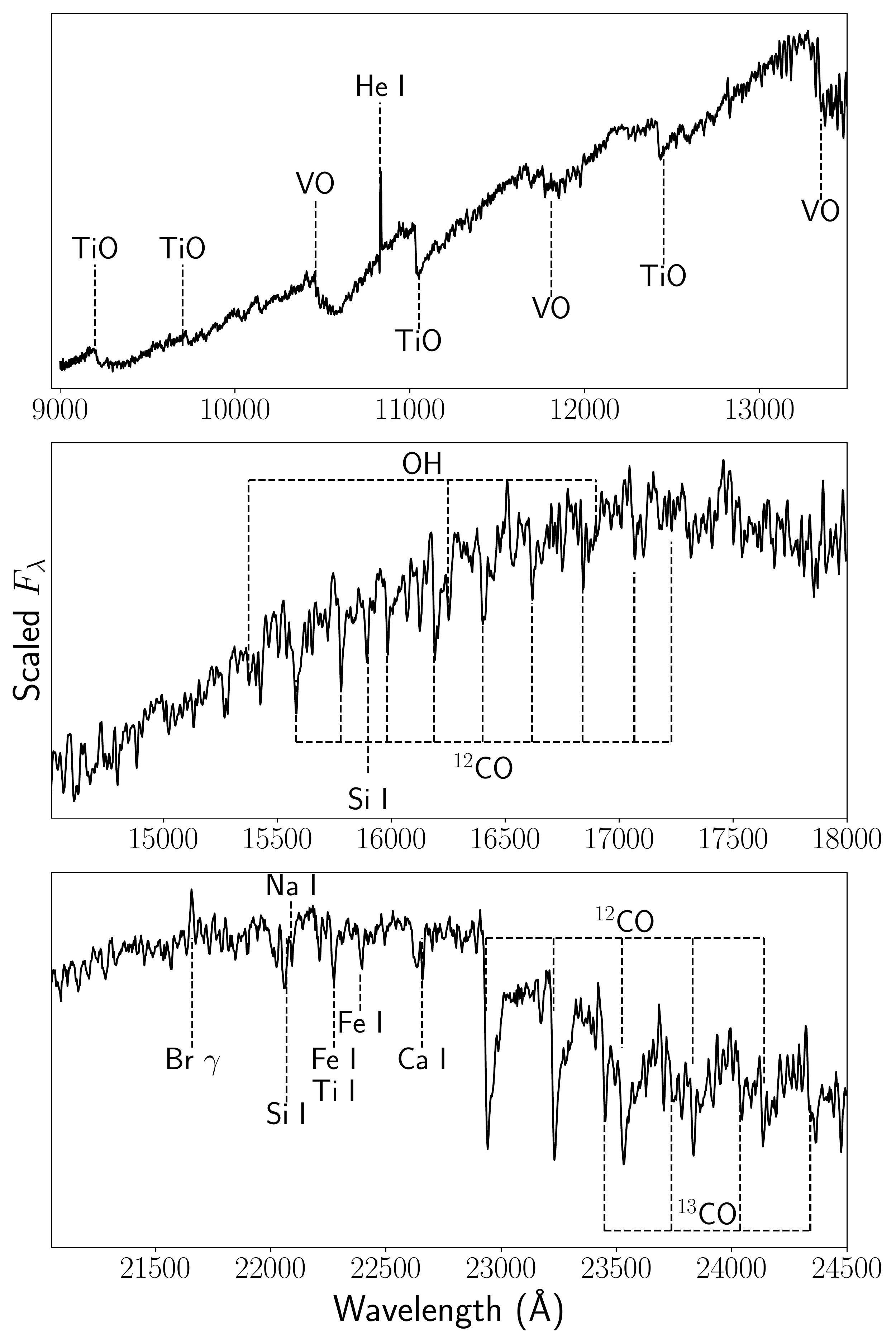}
    \caption{Identification of prominent spectral features in the NIR spectra of IRAS\,18111-2257. The top, middle and bottom panels show spectra in the $J$, $H$ and $K$ bands respectively.}
    \label{fig:nir_ident}
\end{figure}

Figure \ref{fig:nir_ident} shows the NIR spectum of IRAS\,18111-2257 zoomed into the individual $JHK$ bands. The $J$-band spectrum is dominated by absorption bandheads of TiO at 0.97\,$\mu$m, 1.1\,$\mu$m and 1.25$\mu$m, and VO bandheads at 1.046\,$\mu$m and 1.325\,$\mu$m. The TiO bandheads near 1.25\,$\mu$m become discernible at spectral types M7 or later \citep{Wright2009} for cool giants, while the VO features are known to become prominent at very low effective temperatures of $T_{\rm eff} \lesssim 3200$\,K (spectral type M6 or later; \citealt{Hinkle1989, Joyce1998}). The strong TiO band head at 1.1\,$\mu$m detected here is not seen in the spectra of known red supergiants (RSGs) while we also do not detect strong CN features around 1.09$\,\mu$m characteristics of RSGs \citep{Messineo2021}, further arguing against a supergiant classification. The presence of VO instead of the characteristic double absorption bands of ZrO in $J$-band confidently identifies the source as a M-type star instead of a S or C-type AGB star \citep{Hinkle1989}. We identify a strong emission line of He I $\lambda 10833$\,\AA\ on top of the stellar continuum in $J$-band.  We discuss the properties of this feature in Section \ref{sec:hei_wind}.

In $H$-band, we identify Si I 1.59\,$\mu$m, numerous absorption features of the $^{12}$CO second overtone series as well as OH molecular features blended with the CO series. The strong CO second overtone features corroborate the classification of the source as a luminous giant since these features are not seen in M-type dwarfs \citep{Rayner2009}. The most prominent features in $K$-band are the first overtone absorption series of $^{12}$CO and $^{13}$CO. These features are known to be particularly strong in giants and supergiants \citep{Rayner2009}. In addition, we identify strong absorption features from Na I, Si I and Ca I in the $K$-band spectra. While the equivalent widths of these atomic absorption features can be used to constrain the spectral type and luminosity class of the star (e.g. \citealt{Ramirez1997}), the large $K$-band dust excess and water absorption makes this measurement unreliable \citep{Blum2003, Messineo2021}. Overall, the spectroscopic signatures of i) strong water absorption, ii) deep TiO, VO and CO features and iii) weak/absent CN lines suggest the IR counterpart to be a very late-type (M7-M8) O-rich Mira AGB star, akin to the class of D-type symbiotic binaries.

\subsection{Variability}

\begin{figure*}[!ht]
    \centering
    \includegraphics[width=\textwidth]{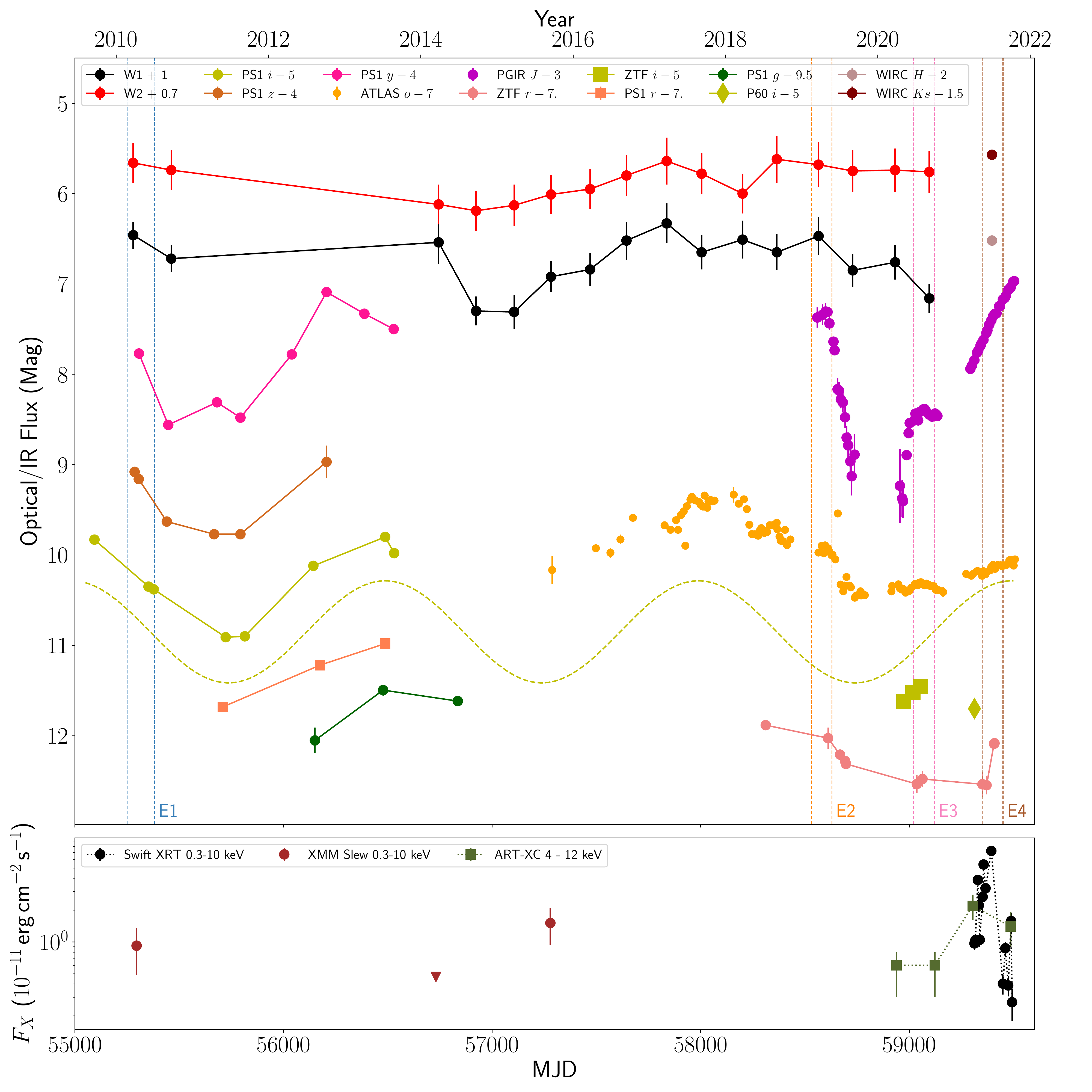}
    \caption{Long term multi-color light curves of IRAS\,18111-2257 (top) along with the long term X-ray evolution combined from archival {\it XMM Newton} data and recent SRG and follow-up data (bottom). For reference, the yellow dashed line in the top panel shows the best-fit 1502 day sinusoidal period overlaid next to the PS1 $i$-band and ATLAS $o$-band data (offset by 0.5 mag from the $i$-band light curve). The dashed vertical lines show the date ranges (and corresponding epoch names) over which the photometric data were combined to perform the multi-epoch dust modeling in Section \ref{sec:dustobs}.}
    \label{fig:hist_lc}
\end{figure*}

Figure \ref{fig:hist_lc} shows the long-term multi-wavelength light curve of IRAS\,18111-2257 over the last $\approx 12$ years. The source exhibits clear variability in from the shortest ($g$-band) to longest ($W2$-band) wavelengths. The $W1$ light curve shows variability with amplitude of $\approx 0.7$\,mag, while the $W2$ light curve exhibits lower variability of $\lesssim 0.5$\,mag. The $J$-band data from PGIR data between 2019 and 2021 shows the largest amplitude variability, with steep drop of $\approx 2$\,mag over $\approx 100$\,days in 2019 followed by a gradual brightening in 2020 and 2021. The source is also systematically fainter in $J$-band (by at least $\approx 2$\,mags) compared to the 2MASS reported magnitudes (from 1998) even accounting for the full variability amplitude observed in PGIR data. The source is also fainter in $H$ and $K$ band when comparing the WIRC and 2MASS magntudes. The ATLAS $o$-band and ZTF $r$-band exhibit smaller amplitude ($\lesssim 0.7$\,mag) drops in 2019 compared to $J$-band.

The PS1 data between 2010-2014 show a long term brightening and fading in all the filters. Given the identification of spectroscopic features similar to Mira variables, we examine the presence of a possible periodicity. Combining the ATLAS $o$-band light curve between MJD 57000 and 58600 (prior to the onset of the abrupt fading episode in 2019) with the PS1 $i$-band light curve, we fit the photometry with a sinusoidal function allowing for a different amplitude between the two filters but requiring a constant period and phase\footnote{The choice of fitting PS1 $i$-band together with ATLAS $o$-band is motivated by the similar wavelength coverage of the two filters as well as to obtain the longest well-sampled baseline between 2010 and 2022. We fix the phase and keep the amplitude variable between the filters since O-rich Mira pulsations exhibit strong amplitude dependence but very weak phase dependence with wavelength \citep{Iwanek2021}.}. We thus derive a period of $1502 \pm 24$\,days. The amplitude is $\gtrsim 1$\,mag in $i$-band, suggesting that the variability is associated with fundamental mode pulsations seen in Mira variables \citep{Wood1996}. However, the likely period lies amongst the longest known in Galactic Mira variables \citep{Soszynski2009, Menzies2019}. We note that the recent optical data in $r$ and $i$ bands from ZTF also show evidence of the same periodicity but are systematically fainter than the corresponding PS1 magnitudes detected $\approx 10$ years ago, even after accounting for the observed variability amplitudes.

\subsection{Distance and Foreground Extinction}
\label{sec:distext}
The spectroscopic diagnostics as well as the likely detection of a $\approx 1500$\,day photometric periodicity suggest that IRAS\,18111-2257 is a long-period pulsating Mira variable. However, the presence of a thick circumstellar dust shell dominating the NIR SED makes the estimation of its luminosity non-trivial from spectroscopic techniques (e.g. \citealt{Ramirez1997}), which is commonly the case for D-type symbiotics \citep{Hinkle2013}. Gaia EDR3 \citep{GaiaEDR3} reports a parallax of $-0.3236 \pm 0.1498$\,mas, while the corresponding distance estimate in the Bailer-Jones catalog \citep{Bailer-Jones2021} is $6.0 - 11.7$\,kpc ($16 - 84$\,percentile confidence). However, Gaia EDR3 also reports a \texttt{astrometric\_gof\_al}\,$\approx 12$ and \texttt{astrometric\_excess\_noise\_sig}\,$\approx 77$, which together with the $>2\sigma$ negative parallax, suggests a very poor (and likely unreliable) astrometric solution.

Next, Mira variables are known to exhibit a period-luminosity (PL) relation \citep{Feast1989, Wood2000} which has been previously used to estimate distances to symbiotic Miras \citep{Whitelock1987, Gromadzki2009}. Using the \citet{Whitelock2008} period-luminosity relation for O-rich variables suggests an intrinsic absolute magnitude of $M_K = -9.9 \pm 0.2$\,mag. However, we caution that the PL relation at such long periods remains debated \citep{Ita2011, Yuan2017} and we do not discuss it further here, but note that many `normal' long period ($> 1000$\,days) O-rich Miras are found to lie on the extrapolation from shorter periods \citep{Whitelock2003}. Using the recent calibration of bolometric luminosity as a function of pulsation period for very long-period O-rich Miras from \citet{Groenewegen2020}, we estimate a bolometric luminosity of $M_{\rm bol} = -6.85 \pm 0.40$ ($L = 4.7^{+2.1}_{-1.5} \times 10^4\,{\rm L}_\odot$).

We then use the period-color relation from \citet{Whitelock2000} to estimate the total (interstellar + circumstellar) extinction, and derive $E(J-K) = 0.31 \pm 0.25$\,mag ($E(B-V) \approx 0.50 \pm 0.40 $\,mag)  based on the 2MASS magnitudes. Based on recent three dimensional dust maps \citep{Bovy2016, Green2019}, the Galactic dust extinction in this direction is estimated to be $E(B-V) \approx 0.9 - 1.3$\,mag for sources farther than $3$\,kpc away. The 2MASS colors are then roughly consistent with the expected interstellar extinction and intrinsic colors of Miras. We thus take the 2MASS measurements to represent the star prior to its current obscured state (see Section \ref{sec:dustobs}). We adopt the foreground extinction to be $E(B-V) = 0.7 \pm 0.6$\,mag, corresponding to $A_K = 0.22 \pm 0.17$\,mag, placing this source at $14.6^{+2.9}_{-2.3}$\,kpc based on the Mira period-luminosity-color relations. The estimated distance and sky location places the source in the Norma arm on the other side of the Galactic center. The total extinction along this line of sight is $A_K \approx 0.5$\,mag \citep{Schlafly2011}, consistent with the source being located within the Galaxy.

\section{A Recent Dust Obscuration Episode}
\label{sec:dustobs}

\subsection{Long term SED evolution}

\begin{figure*}[!ht]
    \centering
    \includegraphics[width=0.8\textwidth]{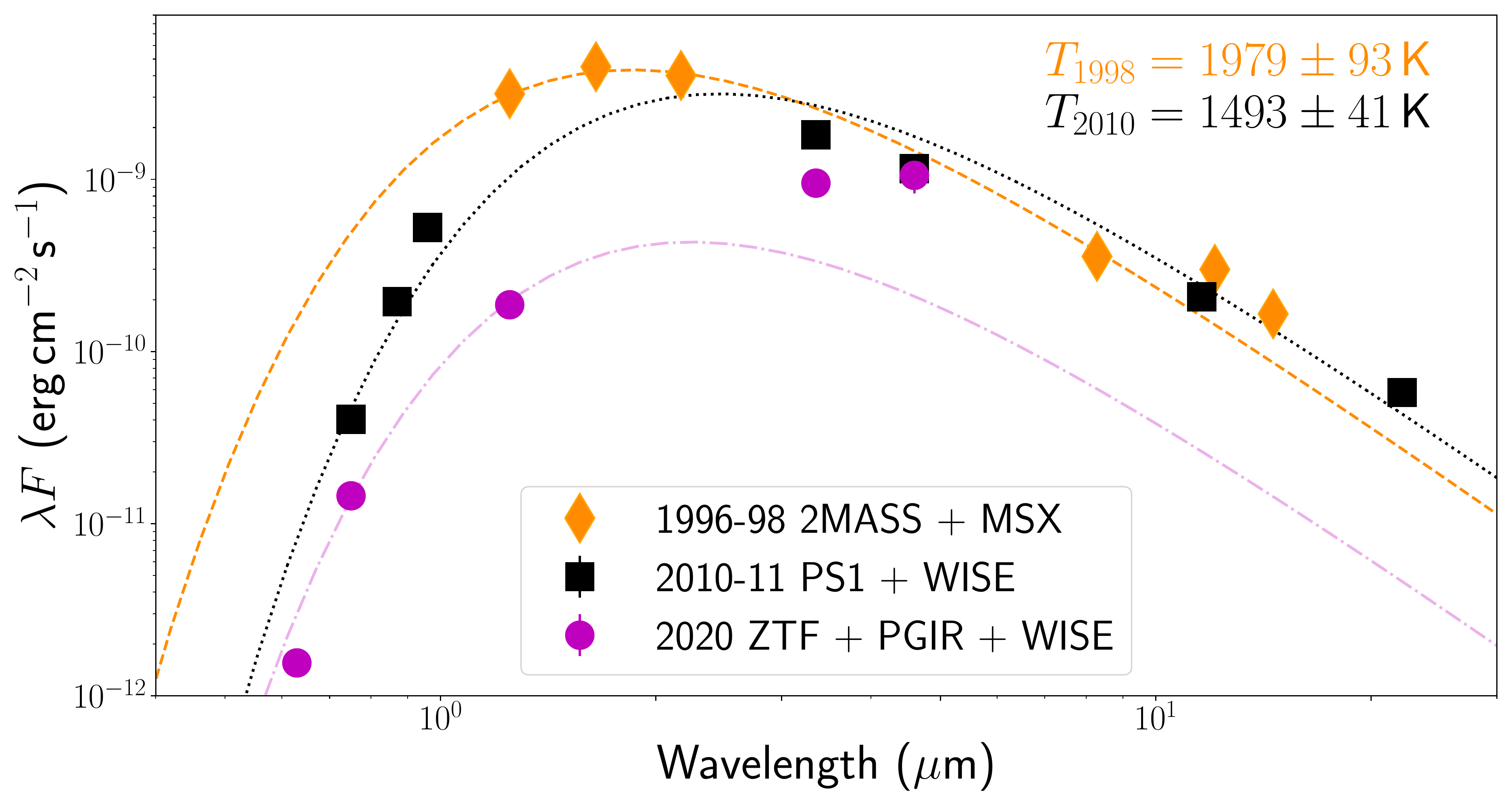}
    \caption{Long term SED evolution of IRAS\,18111-2257. We show the broadband photometric SEDs of the source at three epochs separated by $\approx 10$\,years, corresponding to the epoch of 2MASS (year $\approx 1998$), the first WISE and PS1 surveys (year $\approx 2010$; $E1$ in Figure \ref{fig:hist_lc}) and the last NEOWISE sky survey (year $\approx 2020$; $E3$ in Figure \ref{fig:hist_lc}) respectively. The photometry was corrected for interstellar extinction assuming $E(B-V) \approx 0.7$ (Section \ref{sec:distext}) using the \citet{Cardelli1989} extinction law with $R_V = 3.1$. For each epoch, we also show the best-fit black body curve and the corresponding best-fit temperature. The 2020 SED is not well described by a BB curve and hence we nominally show the best-fit curve without a corresponding temperature. }
    \label{fig:sed}
\end{figure*}

IRAS\,18111-2257 exhibits variability on timescales of $\sim$days to $\sim$years (Figure \ref{fig:hist_lc}). In this section, we analyze the variability of the source and provide evidence for massive dust obscuration episodes over the last $\approx 20$\,years that have progressively obscured the underlying Mira variable. We begin by showing the broad-band photometric SEDs\footnote{We do not use the ATLAS $o$ filter for SED fitting since it has a very wide bandpass covering the $r$ and $i$ bands where the effective extinction varies rapidly. We use $r$ and $i$-band photometry instead, where available.} of the counterpart at three different epochs separated by $\approx 10$\,years in Figure \ref{fig:sed}. The NIR and MIR data for the first epoch capture the peak and Rayleigh-Jeans tail of the SED, which can be well described by a blackbody (BB) of temperature $T \approx 2000$\,K. The apparent BB temperature is consistent with the typical effective temperatures of late type Mira variables at similar $J-K$ color and spectral type \citep{Haniff1995, Feast1996}. Since the 2MASS colors also suggest relatively small circumstellar reddening (Section \ref{sec:distext}), we adopt 2000\,K as the average effective temperature of the underlying unobscured star as observed around the year 1998.

The 2010 SED is characterized by a fainter and colder ($T \approx 1500$\,K) broadband continuum compared to 1998. There are noticable deviations from a simple BB model, which we analyze in Section \ref{sec:dusty}. The 2020 epoch exhibits major differences when compared against both the 1998 and 2010 epochs, characterized by a red SED peaking at $\gtrsim 5\,\mu$m. In addition, the 2020 SED deviates significantly from a single BB model, and we do not attempt to derive a corresponding BB temperature. The source flux has faded $> 10\times$ in $J$-band compared to the 2MASS epoch. Comparing against the long term light curve in Figure \ref{fig:hist_lc}, the dramatic fading was clearly associated with a sudden large amplitude  ($\gtrsim 2$\,mag) drop in the $J$-band flux in late 2019. Such a sudden large dimming is inconsistent with the smooth pulsations in O-rich Miras; however, the observed drop and reddening is strikingly similar to the sporadic dust formation episodes seen in some symbiotic and C-rich Miras (e.g. \citealt{Jurkic2012, Ita2021}).

\subsection{DUSTY modeling}
\label{sec:dusty}
With evidence for recent dust obscuration in IRAS\,18111-2257 based on both the temporal and spectral characteristics, we turn to detailed multi-epoch modeling of the source SED using the radiative transfer code \texttt{DUSTY} \citep{Ivezic1997}. \texttt{DUSTY} solves the radiative transport through a dust shell including dust absorption, scattering and emission assuming spherical symmetry. We model the SEDs at four different epochs listed in Table \ref{tab:dusty} to track the evolution of the mass loss rate and dust parameters of the underlying Mira star. The four epochs were selected to ensure that the wavelength coverage extends from the BB Wien tail of the unobscured star (to constrain the extinction law) to the MIR bands ($\gtrsim 4\,\mu$m; to constrain the dust emission). While \texttt{DUSTY} uses self-similarity and scale invariance to reduce the input parameters, we applied additional constraints based on the inferred stellar parameters.

We model the central star as a BB with $T = 2000$\,K as derived in the previous section, and adopt a distance of $14.6$\,kpc. We assume the dust grains to be composed of warm silicates as in O-rich AGB stars \citep{Ossenkopf1992} with a MRN grain size distribution ($\propto a^{-3.5}$; \citealt{Mathis1977}). We set the minimum grain size to $a_{min} = 0.005\,\mu$m and allow for a variable maximum grain size $a_{max}$. In all cases, we first attempt to fit the SED first with $a_{max} = 0.25\,\mu$m as in standard interstellar grains \citep{Draine2003}, but change $a_{max}$ if no acceptable fits are found. The radial density profile is adopted to be a $r^{-2}$ power law corrected for radiative acceleration in AGB winds \citep{Elitzur2001}. The resulting free parameters of the model are the maximum grain size ($a_{max}$), dust optical depth at $0.55\,\mu$m ($\tau_V$), the shell thickness ($Y$) as a multiplicative factor of the inner radius ($r_{in}$), the dust temperature at the inner edge of the shell ($T_d$) and the luminosity ($L$).

\begin{table*}[]
    \centering
    \begin{tabular}{lccccccc}
    \hline
    \hline
      Epoch  & $L$ & $\dot{M}$ & $a_{max}$ & $\tau$ & $T_d$ & $r_{in}$ & $v_e$  \\
        & $10^4$\,L$_\odot$ & $10^{-6}$\,M$_\odot$\,yr$^{-1}$ & $\mu$m & ($\lambda = 0.55\,\mu$m) & K & AU & km\,s$^{-1}$\\
    \hline
    1996-1998 (E0) & $4.0^{+0.1}_{-0.1}$ & $8.3^{+2.1}_{-1.6}$ & $0.25^{+0.02}_{-0.02}$ & $0.6^{+0.1}_{-0.1}$ & $260^{+40}_{-30}$ & $300^{+115}_{-85}$ & $4.4^{+0.8}_{-0.6}$\\
    2010-2011 (E1) & $2.4^{+0.4}_{-0.2}$ & $6.5^{+0.1}_{-0.7}$ & $0.20^{+0.01}_{-0.01}$ & $7.5^{+0.1}_{-0.1}$ & $1600^{+20}_{-20}$ & $7^{+1}_{-1}$ & $24^{+1}_{-2}$\\
    2019 (E2) & $2.0^{+0.2}_{-0.2}$ & $11.0^{+2.0}_{-1.9}$ & $0.54^{0.13}_{-0.09}$ & $7.0^{+0.3}_{-0.3}$ & $1450^{+170}_{-220}$ & $10^{+2}_{-4}$ & $30^{+8}_{-7}$\\
    2020 (E3) & $1.4^{+0.1}_{-0.2}$ & $21.7^{+3.6}_{-5.6}$ & $2.01^{+0.74}_{-0.29}$ & $5.5^{+0.2}_{-0.1}$ & $760^{+80}_{-80}$ & $37^{+8}_{-11}$ & $14^{+2}_{-2}$\\
    \hline
    \end{tabular}
    \caption{Derived dust parameters from the multi-epoch \texttt{DUSTY} modeling of IRAS\,18111-2257. We list The parameters $\dot{M}$ (the mass loss rate) and $v_e$ (the terminal outflow velocity) are calculated by \texttt{DUSTY} as part of the radiative wind acceleration solution, while other parameters were varied to minimize the least squared residuals from the observed data. The mass loss rates are derived assuming a gas-to-dust mass ratio of 200 and dust grain bulk density of $3$\,g\,cm$^{-3}$.}
    \label{tab:dusty}
\end{table*}

We attempted several different values for the thickness of the shell $Y$ (in the range $\approx 2 - 1000$) but did not find it to significantly affect the derived parameters, and fixed it at $Y = 5$. This is expected since we primarily model the NIR and MIR data ($\lambda \lesssim 10$\,$\mu$m) where the emission is dominated by the hottest and innermost part of the shell \citep{Kotnik-Karuza2007}.  Due to limited historical data, we do not attempt to include the effects of the Mira pulsations in the multi-epoch modeling for simplicity, but note that the stellar effective temperature can change by $\approx 300$\,K \citep{Reid2002}; however the large SED changes modeled here cannot be explained by temperature variations. 

We then used a \texttt{python} wrapper on the \texttt{DUSTY} code to perform least-squares fitting using the Markov Chain Monte Carlo library \texttt{emcee} \citep{Foreman-Mackey2013}. The fitting was performed via minimization of the likelihood function defined using the $\chi^2$ of the model fit and allowing for underestimated uncertainties\footnote{As defined in the \texttt{emcee} online documentation at \url{https://emcee.readthedocs.io/en/stable/tutorials/line/}.}. We assumed flat priors for the fitted parameters and convergence of the posterior sampling chains was ensured \citep{Foreman-Mackey2013}. The best-fit parameters were derived using the median of the posterior sample distribution while their confidence intervals are derived from the 16th-84th percentile (68\% confidence) interval of the distributions. The derived parameters and their uncertainties are listed in Table \ref{tab:dusty} and discussed here.

\begin{figure*}[!ht]
    \centering
    \includegraphics[width=0.49\textwidth]{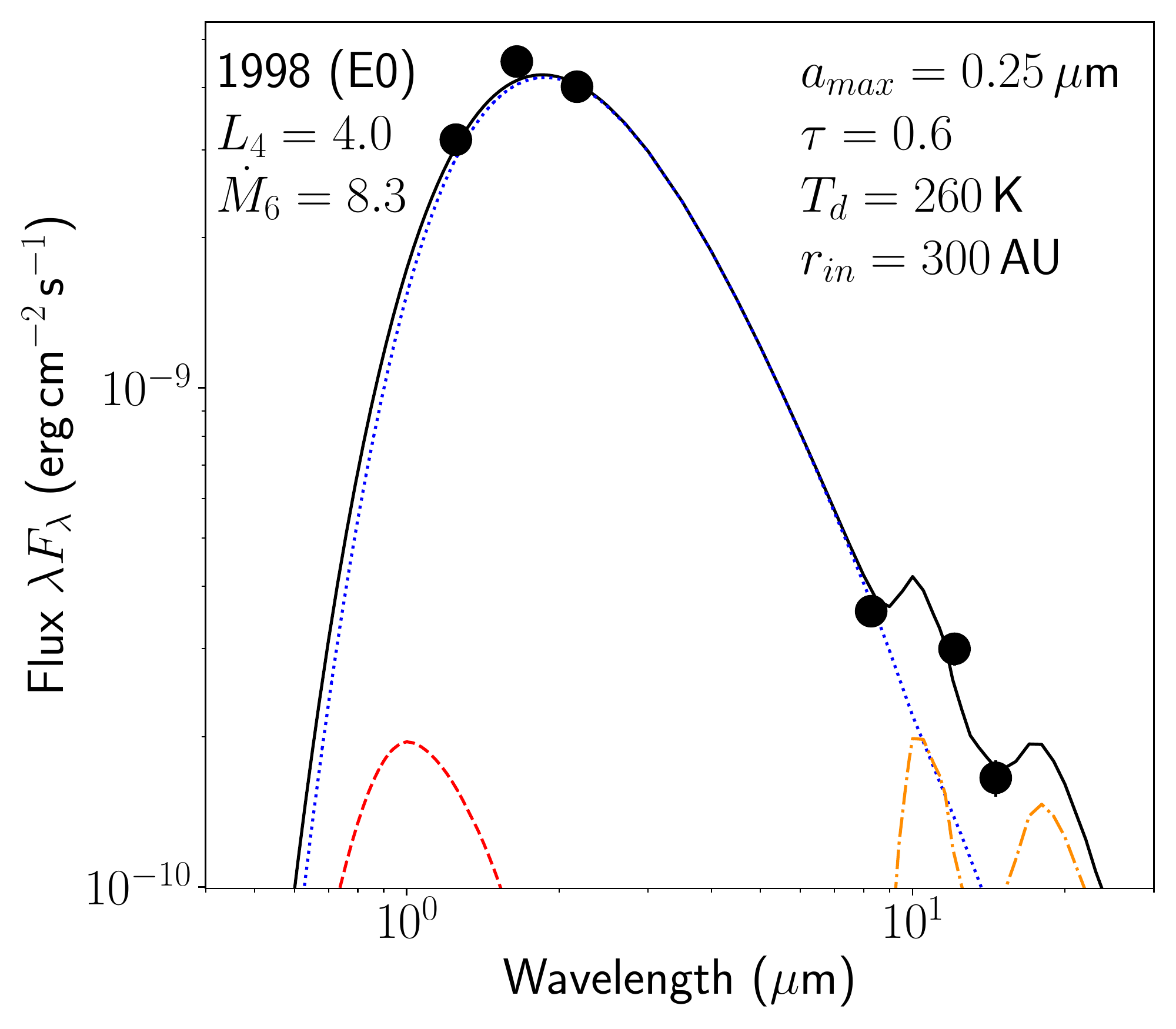}
    \includegraphics[width=0.49\textwidth]{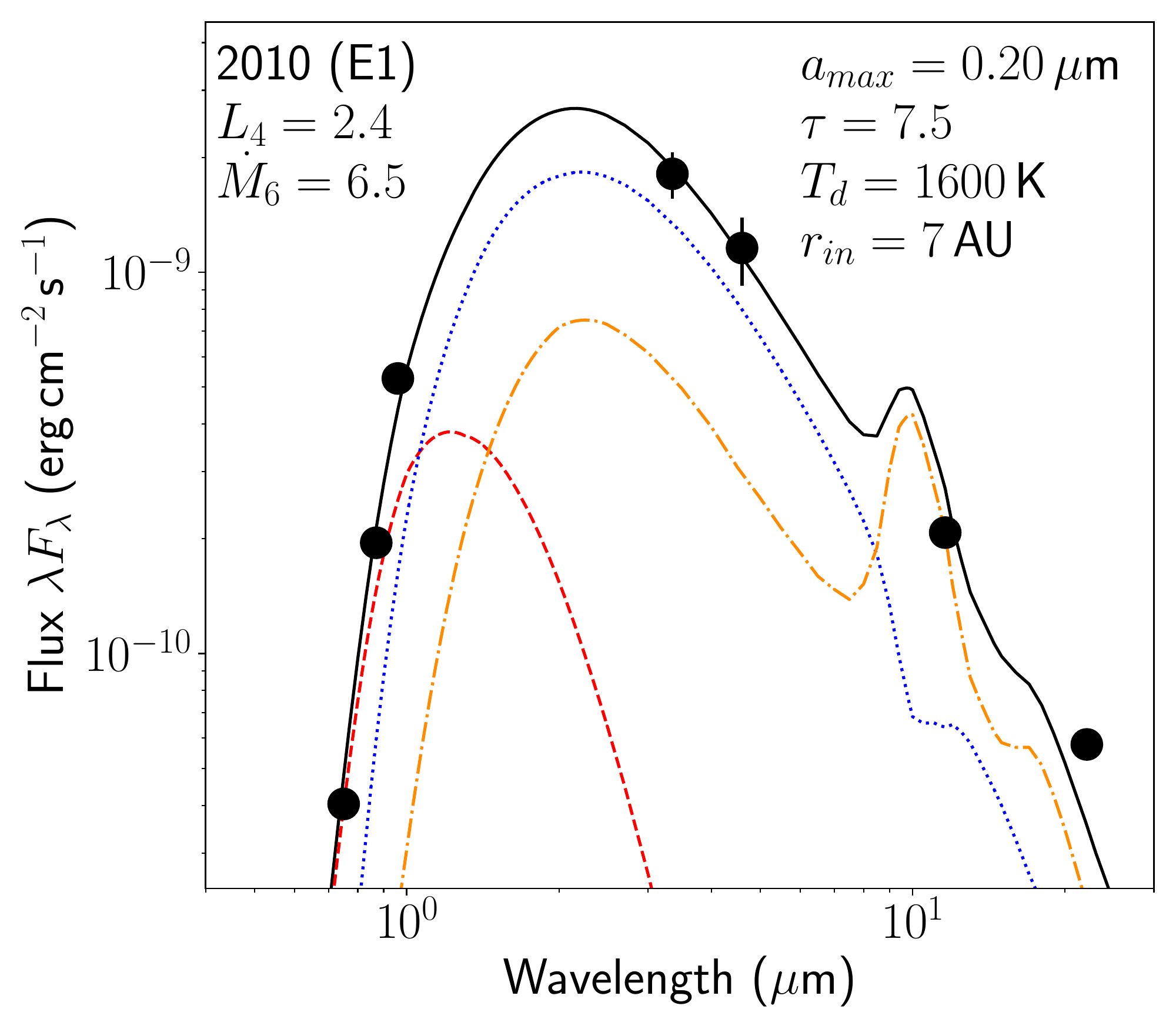}
    \includegraphics[width=0.49\textwidth]{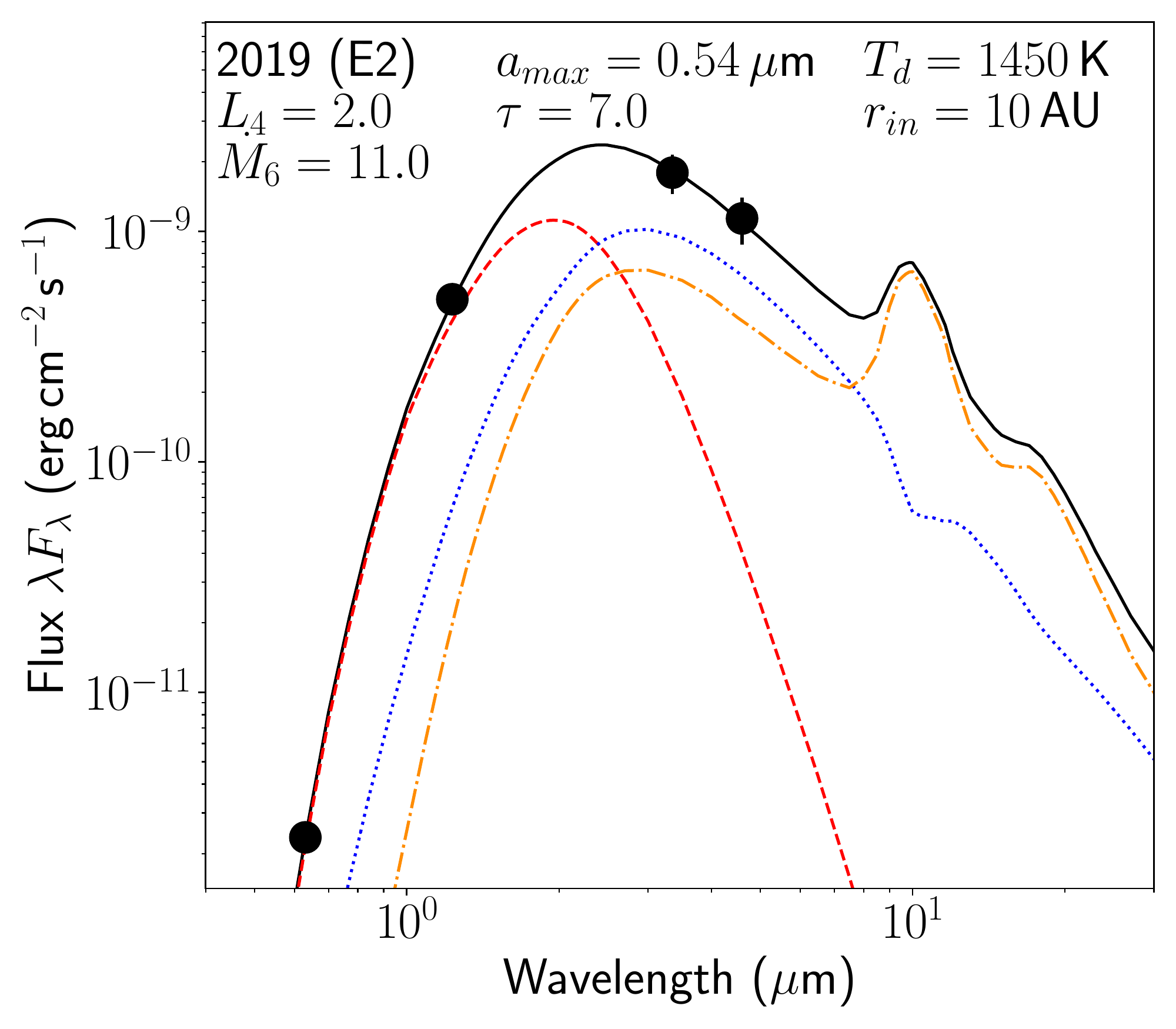}
    \includegraphics[width=0.49\textwidth]{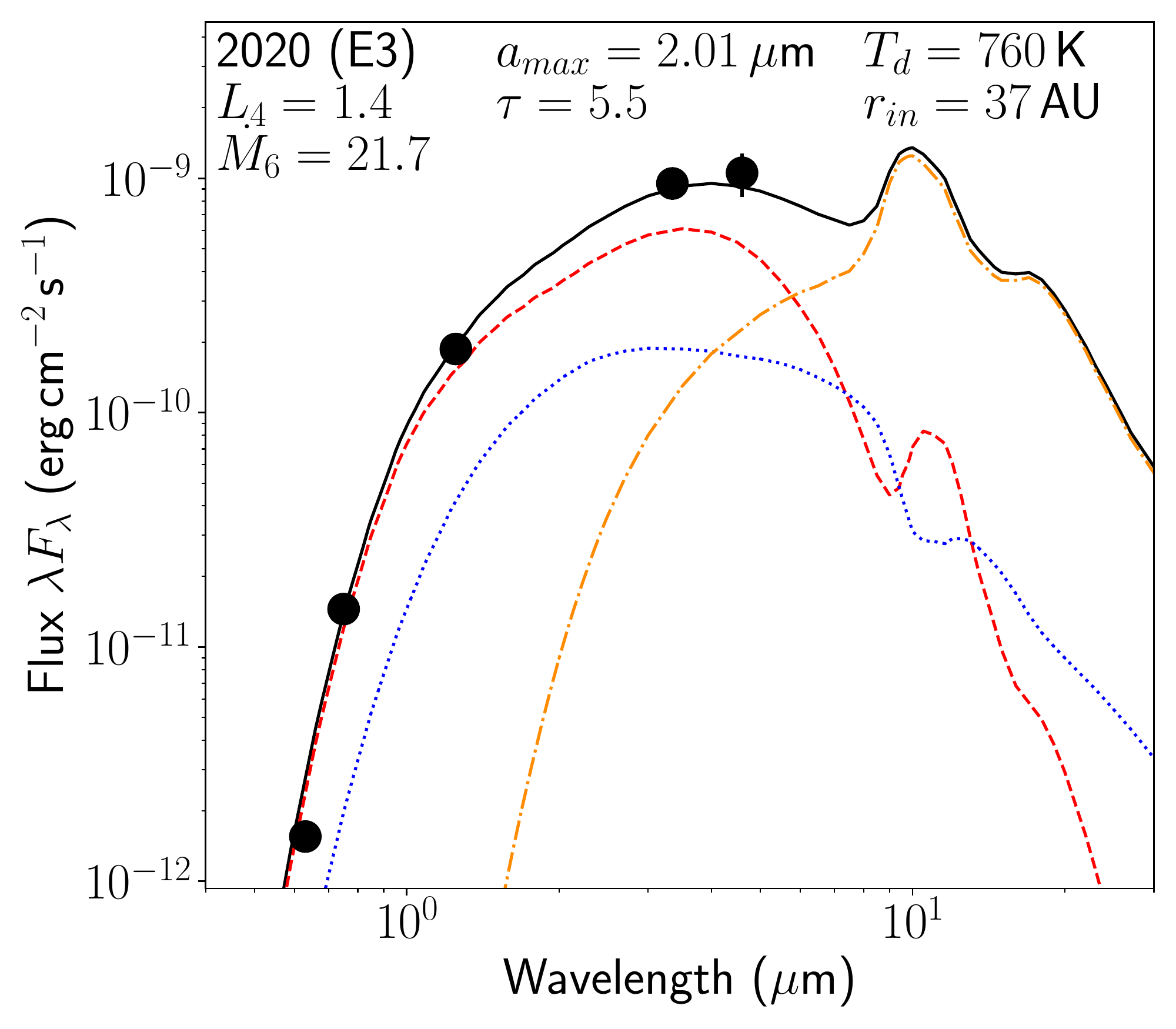}
    \caption{Modeling of the multi-epoch SEDs of IRAS\,18111-2257. In each panel, we show the SED of the source at different epochs (as indicated) along with best-fit models from \texttt{DUSTY} \citep{Ivezic1997}. The black solid lines show the total flux model, the orange dot-dashed lines shows the dust emission, while the red dashed and blue dotted lines show the scattered and attenuated stellar emission respectively. The derived parameters for the model are also indicated in each panel -- $L_4$ indicates the luminosity of the star in units of $10^4$\,L$_\odot$ and $M_6$ denotes the derived mass loss rate in units of $10^{-6}$\,M$_\odot$\,yr$^{-1}$.}
    \label{fig:dusty_seds}
\end{figure*}

\subsubsection{1998: An optically thin, distant and cool dust shell}

Our best-fit model for the observed SED in 1998 is shown in Figure \ref{fig:dusty_seds}. The SED can be well described by a largely unobscured ($\tau \approx 0.6$) stellar BB spectrum with a maximum grain size corresponding to interstellar grains \citep{Mathis1977}. The MSX MIR data show clear evidence for an excess around $\approx 9-11\,\mu$m, corresponding to silicate emission features, consistent with our assumption of the dust composition. The relatively small excess of the feature over the continuum constrains the optical depth to be small, consistent with our assumption of a an unobscured star observed in the NIR bands. The silicate emission feature suggests a cool dust temperature of $\approx 260$\,K located at a distance of $\approx 300$\,AU from the star. The inferred shell parameters are strikingly similar to the cool dust shells detected in D-type symbiotic stars \citep{Angeloni2010, Hinkle2013}.

\subsubsection{2010: Detection of a warm dust shell}

The 2010 SED shows clear deviations from the best-fit model derived for 1998, but can still be explained with nearly interstellar grain size distribution. In particular, the steep drop in flux from NIR to optical wavelengths suggests significant circumstellar extinction and a higher dust optical depth of $\tau \approx 7.6$. The WISE MIR photometry beyond $\approx 10$\,um exhibits clear excess over a simple BB curve (e.g. Figure \ref{fig:sed}) coincident with the silicate and continuum emission from a dust shell. However, the large dust optical depth (from the optical-NIR SED) together with the flux drop from W3 to W4 bands constrains the dust shell to be warm with a temperature of $\approx 1600$\,K, consistent with typical silicate dust sublimation temperatures. Unlike the cool distant dust shell detected in 1998, the warm dust shell is located close to the star at $\approx 10$\,AU suggesting that it may have been recently formed. The slow formation of such warm dust shells is consistent with that seen in many D-type symbiotic stars \citep{Whitelock1983, Shenavrin2011}, and also suggested to be a unique feature of symbiotic Miras \citep{Hinkle2013}. 

\subsubsection{2019: Evidence for progressive obscuration}

The 2019 SED is derived from combining the earliest detections of the source in the PGIR and ZTF surveys together with MIR data from NEOWISE. Compared to the 2010 epoch, the source faded further in the optical $r$-band (Figure \ref{fig:hist_lc}) while the MIR WISE emission remained at a similar level. In particular, combining the ZTF optical data and PGIR NIR data, the models required a change in the maximum dust grain size to simultaneously describe the shallower extinction law and the relatively stable MIR emission. While the optical depth remained similar, we derive a higher mass loss rate suggesting that the star likely underwent increased dust production and mass loss between 2010 and 2019, obscuring the stellar optical-NIR flux as seen in the long term light curves. Evidence for such grain growth under increased mass loss and dust production has also been detected in other symbiotic Miras (e.g. \citealt{Jurkic2010}). 

\subsubsection{2020: A rapid dust obscuration episode}

The 2020 SED data was obtained after the dramatic fading observed in 2019. The observed 2019 fading had the largest amplitude in $J$-band compared to the optical bands while the $W1$ band also exhibited a minor fading. The shallower optical-NIR extinction law uniquely constrains the grain size distribution, requiring it to extend to a larger size of $a_{max} \approx 2.0\,\mu$m. Such large grain sizes were also inferred in \texttt{DUSTY} modeling of the sudden obscuration episodes of the symbiotic Mira RR Tel \citep{Jurkic2012b}. The rising flux in the MIR up to the $W2$ band requires the dust to be cooler ($\approx 750$\,K) than earlier epochs, and located farther away from the star at $\approx 40$\,AU. The change in the dust properties are accompanied by a large increase in the mass loss rate, confirming our suggestion of a dust obscuration episode. The inferred source bolometric luminosity is $\approx 25$\% fainter than that in 2019, which is likely due to the dust formation happening near the minimum of the long term pulsation cycle, as typically seen in other types of pulsating stars \citep{Willson2000}.

\subsection{Grain growth and the peculiar color evolution}

\begin{figure}[!h]
    \centering
    \includegraphics[width=0.49\textwidth]{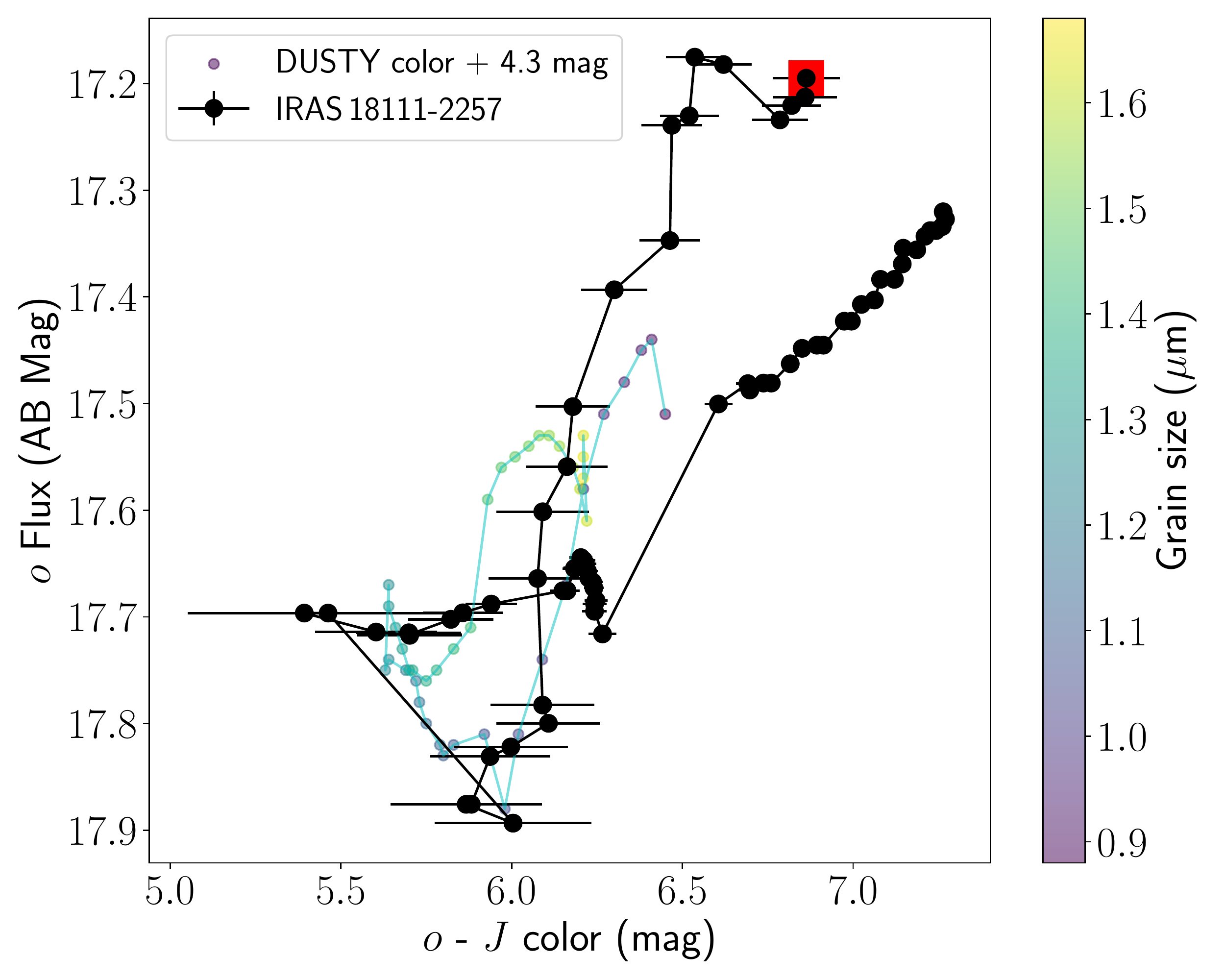}
    \caption{Evolution of the flux and color of IRAS\,18111-2257 during the dust formation episode in 2019. The black dots show the observed color-magnitude evolution as derived from PGIR and ATLAS data. The red square shows the start of the PGIR coverage in 2019. The colored dots show the simulated color evolution as a function of grain size, derived from a grid of models with single-sized dust grains created using \texttt{DUSTY}.}
    \label{fig:dusty_color}
\end{figure}

The $J$-band dimming in IRAS\,18111-2257 during the 2019 dust obscuration episode exhibits a peculiar color evolution. The dimming has a larger amplitude in the NIR $J$-band compared to the optical $r$ and $o$ bands while static dust extinction laws universally suggest larger extinction at shorter wavelengths \citep{Draine2003}. As noted from the \texttt{DUSTY} modeling, the post obscuration SED also require much larger dust grains. The growth of grains during dust formation changes the optical properties of the grains over the duration of the event, suggesting a likely connection with the peculiar color evolution. In order to understand this connection, we show in Figure \ref{fig:dusty_color}, the color magnitude diagram of the 2019-2021 dimming in IRAS\,18111-2257 using PGIR and ATLAS data (which have the highest cadence). 

IRAS\,18111-2257 initially becomes bluer while fading, turns around in a loop during the obscuration minimum and becomes redder while brightening from the dust dip.  Similar color peculiarities have also been discussed in dust formation episodes of the R Coronae Borealis variables \citep{Efimov1988a, Pugach1991}, C-rich Miras \citep{Ita2021} and even noted for the dimming in the O-rich symbiotic nova PU Vul \citep{Efimov1988b}. To understand this phenomenon, we created a grid of \texttt{DUSTY} models to simulate the effect of dust growth on the source colors.  Although we assume a power law grain size distribution for the \texttt{DUSTY} modeling discussed thus far for the individual epochs, the grain extinction evolution may be dominated by specific size ranges during these transient episodes. For simplicity and motivated by the observed evolution of the dust properties between 2019 and 2020 (Table \ref{tab:dusty}), we thus created a model grid varying only the properties of the dust (temperature fixed at $T_d = 1000$\,K) with i) single-sized grains ranging from $\approx 0.5\,\mu$m to $\approx 2.0\,\mu$m and ii) optical depth changing from $\tau = 7.0$ to $\tau = 5.5$.

We perform synthetic photometry on the resulting spectrum grid to simulate the expected color magnitude diagram and also show it in Figure \ref{fig:dusty_color}.  Since in our simple approximation, we do not include the contribution from the static dust distribution prior to the fading (epoch E3; \texttt{DUSTY} does not support multiple grain size distributions), there are differences in the absolute values of the color during the event. However, we find striking similarities in the color-magnitude `loop' observed around the minimum of the dust obscuration event. In reality, the optical depth of the layer, the luminosity of the Mira (which is pulsating) as well as the dust temperature are changing during this event, and hence we do not expect to find perfect agreement between this model and the data.  Nevertheless, as demonstrated in previous works \citep{Efimov1988a, Efimov1988b}, change in dust optical properties during grain growth provides a good explanation for the observed color evolution in the dust formation episode in IRAS\,18111-2257.

\subsection{The He\,I Outflow signature}
\label{sec:hei_wind}

\begin{figure}
    \centering
    \includegraphics[width=\columnwidth]{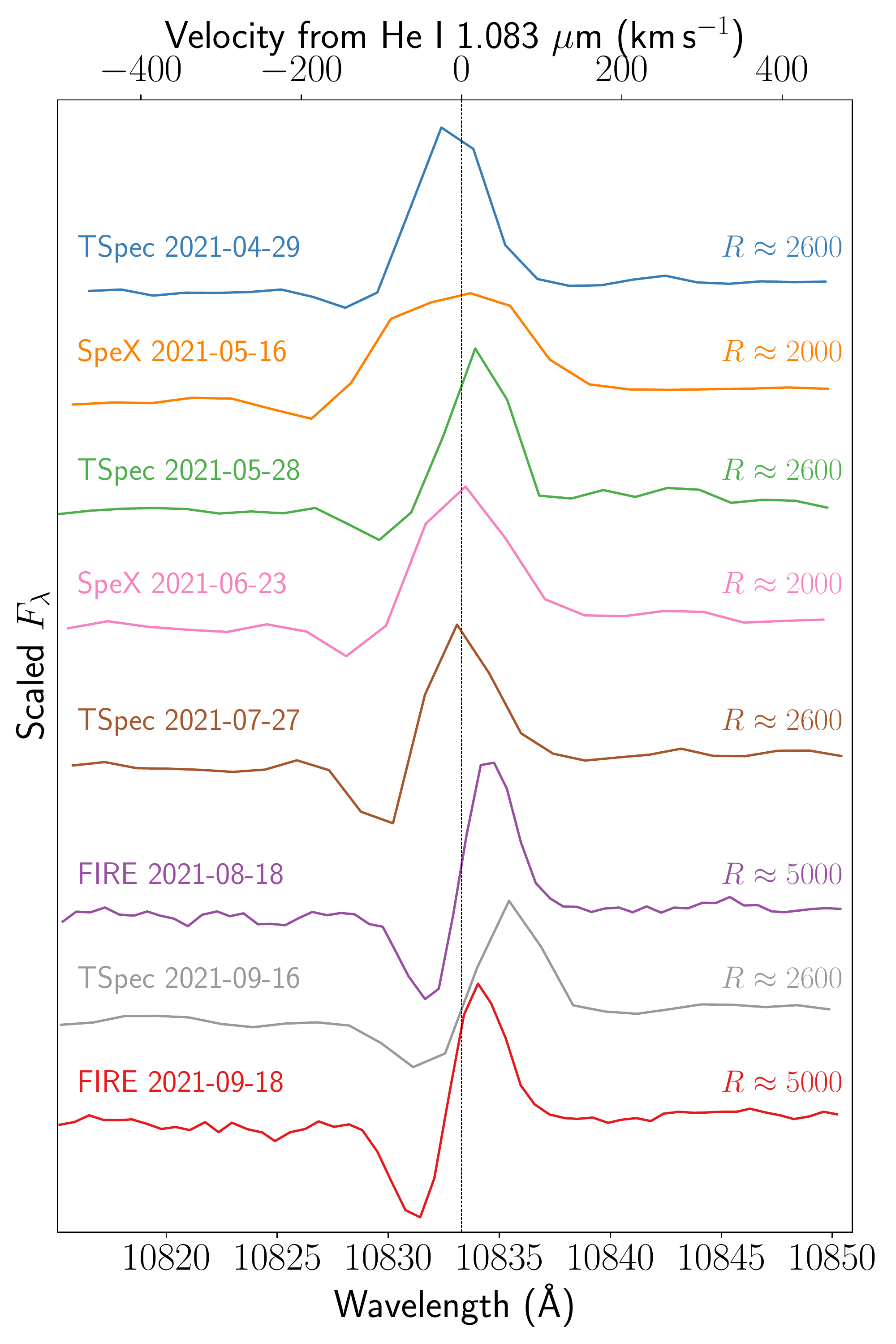}
    \caption{Evolution of the He I 10833\,\AA\ P-Cygni profile in the spectral sequence of IRAS\,18111-2257. We show the NIR spectra zoomed into the region of the line, and indicate the instrument, date and spectral resolution of the observation. }
    \label{fig:hei}
\end{figure}

The NIR spectra of IRAS\,18111-2257 exhibit a strong He I $\lambda 10833$\,\AA\ feature with a P-Cygni profile. Figure \ref{fig:hei} shows the evolution of the line across our spectral sequence. From the highest resolution FIRE spectra, we find that the blue-shifted line absorption extends to velocities of $\approx 100-120$\,km\,s$^{-1}$. \citet{Chakrabarty1998} discovered a similar high velocity ($\approx 250$\,km\,s$^{-1}$) wind in the SyXRB GX\,1+4, although it was unclear if the line arose from the donor star atmosphere or from an accretion disk wind. The He\,I NIR line is detected in many bright giants \citep{Obrien1986}, and with line profiles ranging from pure absorption to P-Cygni shapes \citep{Dupree2009}. The line is formed high in the chromosphere and shown to be a good tracer of bulk motion in the outer atmosphere where the gas flow is close to the escape speed \citep{Dupree1992}.

\citet{Chakrabarty1998} pointed out that the inferred wind velocity in GX\,1+4 was much larger than the typical wind speeds in red giants ($\approx 10-30$\,km\,s$^{-1}$; \citealt{Hofner2018}). They suggested that the UV/X-ray radiation from the hot NS in GX\,1+4 may contribute to accelerating the outflow to high speeds since the line is also detected in other symbiotic stars (e.g. \citealt{Bensammar1988}). \citet{Dupree2009} carried out a survey of the $\lambda 10833$ line in field giants, and found velocities as high as $\approx 90-120$\,km\,s$^{-1}$ in several objects, suggesting that fast outflows may be common. The He\,I $\lambda 10833$ transition lower state is $\approx 20$\,eV above the ground state, and hence it is unlikely to be photoionized by cool photospheres. Indeed, \citet{Dupree2009} showed that the outflows are detected only in stars hotter than $\approx 4500$\,K. With a photospheric temperature of $\lesssim 2500$\,K in IRAS\,18111-2257, the source of the He I ionization, however, remains unclear.

The hot radiation from the compact object may lead to ionization of the donor star's wind, similar to the scenario discussed in \citet{Chakrabarty1998}.  Alternatively, our conclusion of recent dust formation in IRAS\,18111-2257 brings forward another possibility. Dust formation has been extensively monitored in the R Coronae Borealis stars \citep{Clayton1996}, where the He I line is ubiquitously detected only around the epochs of dust formation \citep{Clayton2013}. During these phases, it has been suggested that the outflowing dust cloud shocks the surrounding gas, collisionally excites the transition and radiatively accelerates it outwards to velocities higher than the stellar escape speed \citep{Eyres1999}. The resulting velocities are expected to be $\sim 100$\,km\,s$^{-1}$ for AGB stars with $L \sim 10^4$\,L$_\odot$ \citep{Castor1981}, although optical depth effects reduce the terminal velocities to $\approx 10-30$\,km\,s$^{-1}$ in the steady state winds of AGB stars \citep{Elitzur2001}.

In the case of IRAS\,18111-2257, the derived terminal velocities from the \texttt{DUSTY} solution (Table \ref{tab:dusty}) are $\approx 40$\,km\,s$^{-1}$ around the epochs of dust formation, similar to the observed maximum absorption in the P-Cygni profile. Since the maximum observed absorption speed is $\approx 2-3\times$ higher than the derived terminal velocities, it is possible that the highly variable mass loss rates and rapid grain growth around these epochs accelerates a fraction of the material to higher speeds, as discussed in \citet{Elitzur2001}. Specifically, as the visual optical depth decreases during dust formation (Table \ref{tab:dusty}), the longer wavelength radiation is shifted to shorter wavelengths while the rapid grain growth makes the dust grains further effective at absorbing this emerging radiation and accelerating the surrounding gas with it. We thus suggest that radiative acceleration and collisional excitation during the dust formation episode could also explain the high velocity He\,I outflow observed in IRAS\,18111-2257. In the case of the R Coronae Borealis stars, the line is observed to persist at nearly constant strength and velocity up to $\approx 100$\,days after the obscuration event, since it likely takes time for the far away outer wind to decelerate \citep{Clayton2013}. For the AGB star considered here, the relevant length scales are at least $\approx 10\times$ larger, suggesting that the He\,I line could persist for $\sim 1000$\,days. Continued monitoring of the feature as well as higher resolution spectroscopy to infer its velocity and optical depth evolution (e.g. \citealt{Clayton2003, Dupree2009}) can test the proposed scenario.

\section{Discussion}
\label{sec:discussion}

We have thus far discussed the identification and multi-wavelength characterization of \srgt\, a Galactic hard X-ray transient which exhibits i) a heavily absorbed X-ray spectrum dominated by an optically thick Comptonized plasma at high energies ($> 2$\,keV) and a collisionally ionized plasma at low energies, ii) X-ray variability characterized by `colorless' flares on timescales of $\sim$ minutes and longer term hardening with increasing flux on timescales of $\sim$weeks, iii) a bright infrared counterpart consistent with a luminous pulsating $\approx$M8-type O-rich Mira variable and iv) evidence for an intense dust obscuration episode starting $\approx 800$\,days prior to the peak of the X-ray brightening. In this section, we combine the multi-wavelength properties of the source to constrain the nature of the transient and implications for the broader SyXRB population.

\subsection{Origin of the X-ray emission}

The broadband X-ray spectrum of \srgt\ can be well described by heavily absorbed Comptonized emission with a seed photon temperature of $\approx 1$\,keV and a plasma temperature of $\approx 7.5$\,keV. The spectral shape of the Compton cloud is consistent with a very compact accreting object, likely to be a NS or a BH, surrounded by a corona of hot electrons emitting hard X-rays by Comptonization of soft seed photons. The case for a WD accretor is very unlikely due to the hard X-ray emission extending out to $\gtrsim 30$\,keV as detected with \nustar\ \citep{Luna2013, Masetti2007}. Furthermore, the inferred broadband X-ray flux together with the estimated distance for the IR counterpart ($\approx 15$\,kpc) suggest a peak X-ray luminosity of $\gtrsim 2.5\times10^{36}$\,erg\,s$^{-1}$, much more luminous than accreting WDs ($\lesssim 10^{34}$\,erg\,s$^{-1}$; \citealt{Mukai2017}).

Although we do not detect pulsations or X-ray bursts, the derived spectral parameters are consistent with previously known SyXRBs hosting NSs \citep{Masetti2002, Masetti2007, Masetti2007b, Enoto2014, Kitamura2014}. However, X-ray bursts among SyXRBs are rare and have been detected from only one source (IGR\,J17445-2747; \citealt{Mereminskiy2017, Shaw2020}). In case the accretor is a NS, the seed photon component may arise from a localized hotspot on the surface of the NS, possibly produced by magnetically channelled accretion onto its surface \citep{Masetti2002, Kitamura2014}. We caution that the limited duration for the X-ray follow-up would not be sensitive to the very long spin periods ($\sim $ hours) of some known SyXRBs, and can be masked by the stochastic accretion-related variability (e.g. \citealt{Enoto2014}; Figure \ref{fig:nicerlc}). The pulsed emission may also be washed out by geometric effects or by the bright Comptonizing cloud \citep{Titarchuk2002}.

The X-ray temporal and spectral variability offers additional insights into the accretion process. The transient brightened for $\approx 100$\,days after discovery and has nearly faded back to its quiescent level seen prior to the ART-XC discovery. The source became harder during its brightening phase as typically seen in other SyXRBs, and is consistent with accretion from a wind with a low specific angular momentum \citep{Smith2002, Wu2002}. The variable column density and covering fraction of the partial absorber in our spectral modeling across the epochs (Table \ref{tab:nimodel_evol}) suggest accretion inhomogeneties affecting the spectrum on timescales of $\approx$weeks. On the other hand, the spherically symmetric \texttt{Tbabs} component exhibits relatively constant $N_{\rm H}$ across all the epochs, suggesting that it is associated with a constant contribution from circumbinary material. For the estimated optical depth of the dust shell around the donor in the most recent epoch ($\tau_V \approx 5$), the corresponding $A_V \approx 5.5$\,mag implies a hydrogen column density of $\approx 1.2 \times 10^{22}$\,cm$^{-2}$\citep{Guver2009}, consistent with the \texttt{Tbabs} component. The \texttt{pcfabs} component exhibits a $\approx 3\times$ larger column density, suggesting enhanced absorption close to the compact object.

We consider two alternative possibilities to explain the short-term variability ($\sim $minutes). Since the hardness ratio remains nearly constant, the short-term variability cannot be attributed to a variable covering fraction or column density, since we would expect spectral shape changes as the estimated column from our spectral modeling is too low to be opaque to X-ray photons at $\approx 1-10$\,keV. Thus, we speculate that variable accretion at constant covering fraction and column density arising from the donor wind may explain the observed variations. Alternatively, we suggest that these variations may be attributed to changing Comptonization of the thermal seed photons. Specifically, even if the thermal seed component remains roughly constant, a variable Comptonization fraction at a constant plasma temperature can produce changes in the integrated luminosity of the \texttt{comptt} component with only minor changes to its spectral shape. Since the spectral modeling demonstrates that the optical depth is high and the spectrum is dominated by the Comptonized emission at $\gtrsim 1$\,keV, the observed achromatic variations may reflect the changing luminosity and constant spectral shape of the \texttt{comptt} component.

\subsection{A dense ionized nebula around the compact object}

The broadband spectral analysis reveals a soft flux excess (below $\approx 1$\,keV) that we model with an optically thin collisionally ionized plasma (\texttt{apec}) component. The model is motivated by the detection of strong emission lines coincident with transitions of Mg and Si in the \nicer\ spectra. The derived fits suggest a plasma temperature of $\approx 0.6$\,keV and indicate that this component is located outside the dense partially absorbing cloud, and thus relatively extended compared to the central X-ray emitting region around the compact object. Collisionally ionized plasmas are commonly observed in the X-ray spectra of dense shocks in colliding wind binaries and supernovae (e.g. \citealt{Oskinova2017, Chakraborti2012}). Taking the estimated distance of the counterpart, the derived normalization of the \texttt{apec} component suggests an emission measure of $\approx 2 \times 10^{57}$\,cm$^{-3}$. If we take the relevant size for the extended \texttt{apec} emission region to be of the order of the inferred dust shell ($\sim 10$\,AU), we derive a required particle density of $\sim 10^7$\,cm$^{-3}$ assuming $n_H \approx n_e$. The effective column density arising from the \texttt{apec} component is then $\sim 10\,{\rm AU} \times 10^7\,{\rm cm}^{-3} \sim 10^{21}\,{\rm cm}^{-2}$, and much smaller than the cold absorber column. In Section \ref{sec:acc_nature}, we show that the derived particle density is consistent with an ionized nebula formed from ejected material from the donor star.

Supporting the presence of a surrounding gas cloud, the X-ray spectra show multiple narrow emission and absorption features. We detect prominent and variable narrow Fe K$\alpha$ emission in all the epochs, characteristic of wind-fed systems where the line is produced by X-ray fluorescence of the companion wind material. We do not find evidence for spectral residuals suggestive of cyclotron lines seen in some accreting pulsars \citep{Staubert2019} and possibly even in a SyXRB \citep{Bozzo2018}. A striking feature noted in the \nicer\ spectral analysis is the presence of a strong Si K-edge that we modeled using an additional \texttt{edge} component. The improvement in the spectral fit suggests that the gas surrounding the compact object contains an overabundance of Si\footnote{Since interstellar Si abundances are already included in the \texttt{Tbabs} and \texttt{pcfabs} components, improvements in the fit after including additional edges suggests an overabundance}, which is known to be depleted into silicate dust grains in the interstellar (ISM) medium \citep{Draine2003, Schulz2016}. The edge exhibits a variable optical depth, as visually apparent in the multi-epoch spectra (Figure \ref{fig:specevol}).

Following \citet{Schulz2016}, the measured (excess) optical depth of the Si edge (relative to ISM) can be related to the total Si abundance using,
\begin{equation}
    \tau_{Si} = N_{\rm H} A_z (x-1) \sigma_z
\end{equation}
where $\tau_{Si}$ is the optical depth of the edge, $A_z$ is the expected abundance (per hydrogen atom) in the ISM, $x$ is the total Si abundance relative ISM composition and $\sigma_z$ is the absorption cross-section. Using the assumed \citet{Wilms2000} ISM abundances, $\tau_{Si} \approx 0.3$ measured from the brightest \nicer\ spectrum and a corresponding total $N_{\rm H} \approx 5 \times 10^{22}$\,cm$^{-2}$ (Table \ref{tab:nimodel_evol}), we find that the material surrounding the accretor contains $\approx 2.5\times$ Si abundance relative to normal ISM. However, we caution that scattering effects are known to alter the simple absorption edge profiles used in \texttt{xspec} \citep{Corrales2016}, which can affect the inferred abundance \citep{Schulz2016}; such effects are only probed with higher resolution spectroscopy. Nevertheless, we conclude that multiple diagnostics in the X-ray spectra securely demonstrate the existence of a dense dust/gas cloud enshrouding the binary system.

\subsection{A Mira donor star exhibiting intense mass loss}

The classification of the donor star as a very late-type pulsating AGB star offers a novel addition to the growing population of Galactic SyXRBs. The majority of previously known SyXRBs (Table 1 in \citealt{Yungelson2019}) contain early M-type donors consistent with red giants, in addition to some sources where the donor has been suggested to be a red supergiant \citep{Hinkle2020, Gottlieb2020}. The brightest known source GX\,1+4 also has a M6-III red giant donor that is nearly overflowing its Roche Lobe and powering the luminous X-ray emission \citep{Chakrabarty1997}.  The classification of the donor in the SyXRB Scutum\,X-1 was recently revised from a red supergiant \citep{Kaplan2007} to a long-period Mira variable \citep{De2022} similar to \srgt. XTE\,J1743-363 was also suggested to host a $>$M7 type donor star, and likely a pulsating variable on the AGB \citep{Smith2012}. In the X-ray band, both Sct\,X-1 and XTE\,J1743-363 exhibit similar behavior as \srgt\, including $\sim 100$\,day long flares (\citealt{Smith2012}) as well as variability on timescales of $\sim$hours \citep{Sguera2006, Kaplan2007}. However, the donor photometric variability in XTE\,J1743-363 remains unconstrained.

The pulsation period of $\approx 1500$\,days and O-rich composition provides strong constrains on the evolutionary stage of the donor star and binary system. Such long period Miras are rare, likely because they are in a very short lived evolutionary phase \citep{Marigo2017} or they are realized only among the rarer massive AGB stars \citep{Feast2009}. Indeed, there are $< 20$ candidate Miras in the Milky Way with periods $> 1000$\,d \citep{Menzies2019}, confirming their rarity. However, Figure \ref{fig:srg_lc} shows that our combined photometric dataset over the last $\approx 12$ years supports the presence of a coherent $\approx 1502$\,day period already covering $\approx 3$ cycles of the Mira pulsation. As in \citet{Groenenwegen2018}, we employ the relationship derived in \citet{Groenenwegen2017} based on Cepheid models \citep{Bono2000} between the luminosity ($\approx 2 \times 10^4$\,L$_\odot$), temperature ($\approx 2000$\,K), metallicity (assuming $Z\approx 0.015$) and mass for fundamental mode pulsators to estimate a current mass of $\approx 5.2$\,M$_\odot$ for the Mira variable, making it a massive AGB star. The simpler period-mass-radius relationship in \citet{Wood1990} provides a similar estimate of $\approx 5.5 - 8.0$\,M$_\odot$ depending on the adopted radius\footnote{We estimate the radius to be $\approx (8-10) \times 10^{13}$\,cm, based on the estimated bolometric luminosity near 2019 and the assumed stellar temperature.}.

Thanks to multi-epoch archival photometric data as well as high cadence monitoring near the X-ray flare, we present strong evidence for an intense dust obscuration episode in IRAS\,18111-2257 starting $\approx 800$\,days prior to the observed X-ray brightening. In addition, the long $\approx 25$\,year baseline between 2MASS, PS1, WISE and PGIR allows us to observe a dramatic transition in the appearance of the donor star from a relatively unobscured stellar SED (i.e. an S-type system) to a heavily dust obscured Mira (i.e. a D-type system). The secular change is accompanied by an increase in the mass loss rate (indicated by the \texttt{DUSTY} modelling) and suggests that the donor star is possibly at the onset of a thermal pulse \citep{Herwig2005}. The long pulsation period and high inferred mass points to a relatively young system with age $\lesssim 100$\,Myr \citep{Marigo2017}; comparing the inferred current mass to the grid of intermediate mass AGB models in \citet{Doherty2015}, we estimate an initial mass of $\approx 6$\,M$_\odot$ and age of $\approx 90$\,Myr if the donor is indeed entering the first thermal pulse.

\subsection{The nature of the accretion flare and binary system}
\label{sec:acc_nature}

The coincidence of the dust formation (Section \ref{sec:dusty}) with the observed X-ray flare in 2021 suggests that the enhanced mass loss provides additional fuel for accretion onto the compact object, that likely caused the X-ray brightening. This allows us for the first time to constrain the binary orbit in a SyXRB together with the accretion mechanism by comparing the properties of the mass loss episode and the observed X-ray flare. The \swift\ monitoring shows that the X-ray outburst reached maximum luminosity at $t_d \approx 800$\,days from the start of the dust obscuration episode in 2019. If we take the He\,I P-Cygni signature as an estimate of the terminal wind velocity ($v_t \approx 100$\,km\,s$^{-1}$, accelerated by the dust production or X-ray emission; Section \ref{sec:hei_wind}), the compact object is likely located $ a \sim v_t \times t_d \approx 55$\,AU from the donor star.

\begin{figure*}[!ht]
    \centering
    \includegraphics[width=\textwidth]{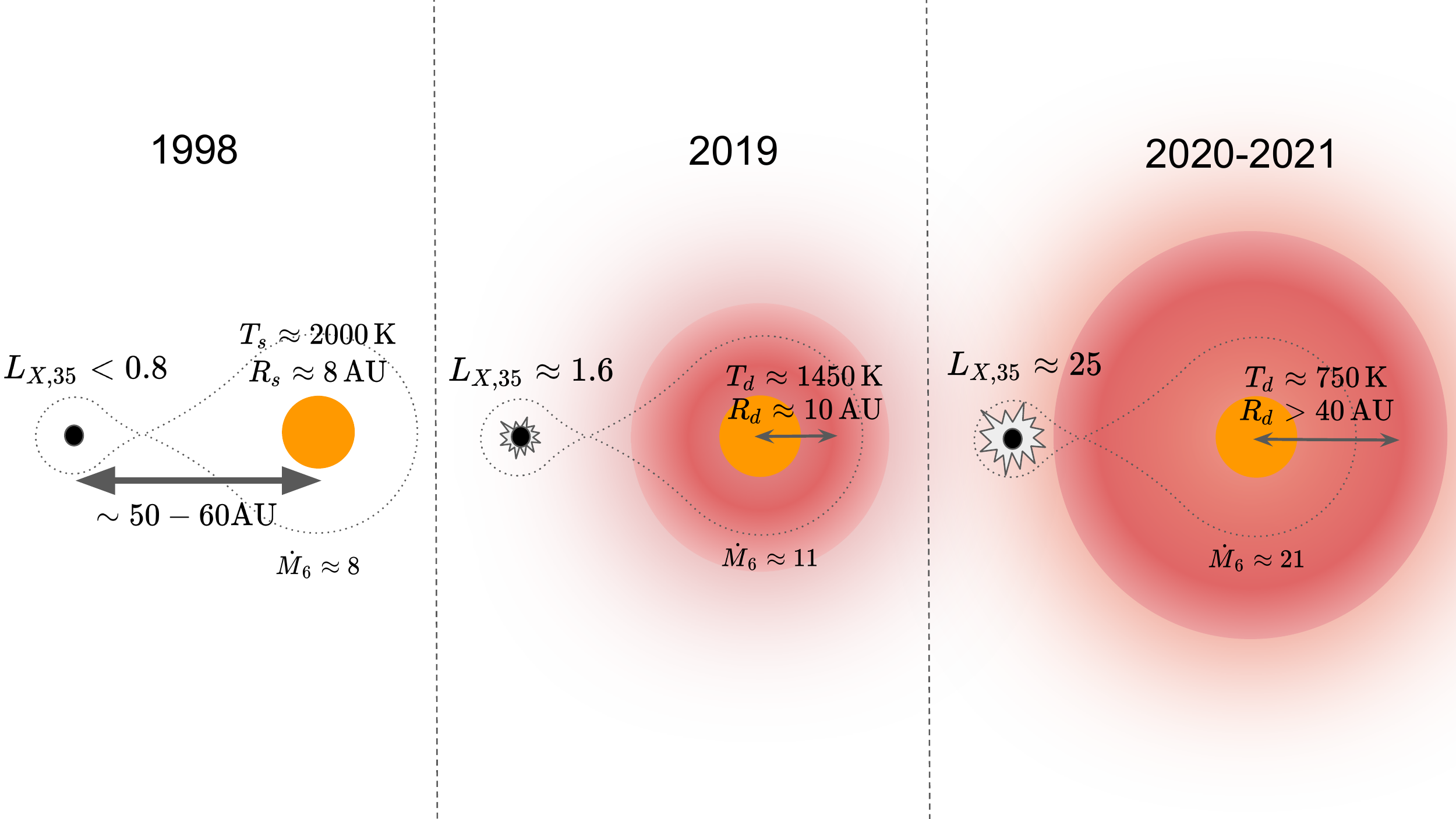}
    \caption{Proposed model for the enhanced mass loss and X-ray activity in \srgt. We show depictions of the binary system for three epochs, as inferred from our modeling. The left configuration shows the expected state of the binary near the epoch of 2MASS, consisting of an unobscured late-type giant with a temperature $T_s$, radius $R_s$ and mass loss rate $\dot{M}_6$ (in units of $10^{-6}$\,M$_\odot$\,yr$^{-1}$). The donor star fits well inside its Roche lobe (shown by the dotted lines) as it orbits a distant compact object that has a faint X-ray accretion luminosity $L_{X, 35}$ (in units of $10^{35}$\,erg\,s$^{-1}$) as constrained by {\it INTEGRAL} archival hard X-ray data. In the middle configuration, the compact object is marginally brighter in the X-ray around the epoch of the first ART-XC survey, as the donor star shows evidence for elevation in mass loss rate and the progressive formation of a warm dust shell with an inner radius $R_d$ and temperature $T_d$. The shell has a thickness ratio of $\approx 5$, and we only show the inner dense region here for clarity. In the right configuration, the donor undergoes a dramatic dust dimming episode in 2019-2020 as it ejects an expanding gas/dust shell that slowly expands beyond the Roche lobe, cools and reaches the compact object. The enhanced gas supply causes a $\approx 15\times$ X-ray flux brightening on the compact object, while also enshrouding it in gas and silicate dust directly detected in the X-ray spectra.}
    \label{fig:srgt_model}
\end{figure*}

Assuming the mass ratio ($q$) of the binary system to be $\approx 4$, assuming the compact object to be a $\approx 1.3$\,M$_\odot$ NS\footnote{The constraints on the binary configuration and mass accretion process do not change substantially if we assume a $10$\,M$_\odot$ BH.} and the donor to be a $5$\,M$_\odot$ AGB star, the Roche Lobe radius for the donor would be $\approx 23$\,AU, larger than the estimated photospheric radius ($\lesssim 6$\,AU) of the unobscured star, suggesting that the system would be nominally undergoing wind accretion. However, our \texttt{DUSTY} modeling shows that the intense dust obscuration episode in 2020 results in the formation of an expanding dust envelope. The densest inner region of the dust shell is observed to expand from $\approx 10$\,AU in 2019 to $\approx 40$\,AU in 2020, thus nearly reaching the location of the compact object shortly prior to the start of the X-ray brightening.

For the assumed shell thickness ($Y=5$), the entire binary system is likely already engulfed in the dust shell prior to the X-ray brightening, consistent with the strong Si over-abundance observed in the X-ray spectra. We can constrain the particle density in this region by approximating the wind density profile with standard $r^{-2}$ dependence at large distances from the star, as we also confirmed from the radial density profile produced from the \texttt{DUSTY} model. Then, the mass loss rate is related to the radial density profile as
\begin{equation}
    \rho = \frac{\dot{M}}{4 \pi v_t} r^{-2}
\end{equation}
For a distance of $\approx 50$\,AU from the donor to the compact object, we estimate a gas particle number density of $\approx 1.1 \times 10^7$\,cm$^{-3}$. This density is similar to what we estimate for the low energy \texttt{apec} component observed in the X-ray spectra, suggesting that ionization of the dense Mira wind near the compact object provides a good explanation for the extended collisionally ionized plasma detected around the compact object.

For the low angular momentum dusty wind of the donor star, the accretion is expected to take place directly onto the compact object without the formation of a disk \citep{Smith2002, Wu2002}. Following \citet{Yungelson2019}, we can estimate the expected accretion rate assuming the Bondi-Hoyle-Lyttleton formalism. The compact object captures material within a characteristic Bondi radius given by
\begin{equation}
R_B \approx \frac{2 G M_{c}}{ v_{\rm rel}^2 + c_s^2}
\end{equation}
where $R_B$ is the Bondi radius, $M_{c} \approx 1.3$\,M$_\odot$ is the mass of the compact object assumed to be a NS, $c_s$ is the sound speed, $v_{\rm rel} = \sqrt{v_{\rm orb}^2 + v_w^2}$, $v_{\rm orb}$ is the relative orbital velocity and $v_w$ is the wind velocity. For the wide apparent orbit and cool wind considered here, we ignore the NS orbital velocity and sound speed of the gas relative to the much faster wind velocity, resulting in $R_B \approx 0.25$\,AU. Then the accretion rate onto the compact object is
\begin{equation}
    \dot{M_B} = \frac{1}{4} \dot{M_d} \frac{v_{\rm rel}}{v_w} \left(\frac{R_B}{a}\right)^2
\end{equation}
where $\dot{M_B}$ is the Bondi accretion rate and $\dot{M_d}$ the donor mass loss rate. Taking the enhanced mass loss rate from the donor estimated from the 2020 epoch ($\dot{M_d} \approx 2.2 \times 10^{-5}$\,M$_\odot$\,yr$^{-1}$), we derive an accretion rate of $\dot{M_B} \approx 1.1 \times 10^{-10}$\,M$_\odot$\,yr$^{-1}$. Assuming a fraction $\epsilon$ of the rest mass energy of this material is emitted by the compact object during the X-ray flare peak ($L_X \approx 2.5 \times 10^{36}$\,erg\,s$^{-1}$), we estimate $\epsilon \approx 0.3$, similar to the expected conversion efficiency for accretion onto a NS ($\epsilon \sim 0.2$). A lower conversion efficiency would be required if the ejected dust shell is focused by the Roche potential into the plane of the binary system \citep{Shagatova2021}. If we assume that the wind velocity prior to dust formation was close to that estimated by \texttt{DUSTY} prior to the onset of dust formation (E1; 2010-11) for a steady state wind (Table \ref{tab:dusty}), the $\approx 4 \times$ lower mass loss rate would both reduce the mass transfer rate onto the NS as well as reduce the accretion efficiency due to the larger magnetospheric radius at low accretion rates \citep{Kuranov2015}. We summarize the proposed scenario for the X-ray flare in \srgt\ in Figure \ref{fig:srgt_model}.

\section{Summary}
\label{sec:summary}

We have presented the discovery and detailed multi-wavelength characterization of the Galactic hard X-ray transient \srgt\ discovered by the \artxc\ instrument on board the SRG satellite. In summary, we show that
\begin{itemize}
    \item The X-ray transient exhibited a $\approx 100$\,day brightening to peak following the discovery (as observed with \swift), reaching a peak luminosity of $\approx 2.5 \times 10^{36}$\,erg\,s$^{-1}$, followed by fading nearly to its quiescent ($\approx 10\times$ fainter) flux level. In addition, the source is observed to exhibit short timescale `colorless' flares changing by $\approx 5\times$ in flux over timescales of $\lesssim 1$\,hour.
    \item Modeling of the broadband X-ray spectroscopy from \nicer\ and \nustar\ reveals a highly absorbed spectrum with emission extending out to $\gtrsim 30$\,keV. The properties are consistent with Comptonization of low energy seed photons by a hot plasma around a NS or BH accretor. The source spectrum becomes harder as it brightens during the X-ray flare.
    \item We do not find any evidence for pulsations or bursts that would confidently suggest the presence of a NS accretor. We detect the presence of a low energy excess exhibiting likely emission lines of Mg and Si, which we model as an optically thin collisionally ionized plasma with a temperature of $\approx 0.6$\,keV.
    \item Imprinted on the broadband X-ray spectrum, we detect narrow and variable Fe K$\alpha$ emission likely arising from X-ray fluorescence of relatively distant material. We also identify an excess of Si K-edge absorption, which we interpret as an overabundance of silicate dust around the system.
    \item Spatially coincident with the X-ray transient, we identify a bright and heavily obscured IR source, IRAS\,18111-2257, which we demonstrate to show correlated temporal variability using data from the Palomar Gattini-IR survey.
    \item With optical/IR spectroscopy of the counterpart, we classify the star to be a very late-type (M7-M8), pulsating ($P\approx 1500$\,days) and luminous ($M_K\approx -10$) Mira variable on the AGB, located at a distance of $\approx 15$\,kpc. The X-ray properties and the IR identification suggest that \srgt\ is a new Galactic symbiotic X-ray binary.
    \item Collating optical/IR photometric data from the last $\approx 25$\,years, we show that the IR source has undergone dramatic reddening likely associated with enhanced mass loss and dust production. In particular, the PGIR $J$-band data conclusively show evidence of an intense fading episode of the counterpart in 2019, $\approx 800$\,days prior to the X-ray flare.
    \item We model the multi-epoch optical to mid-IR SEDs of the source with \texttt{DUSTY} to show that the dramatic fading observed in the NIR can be explained by an intense dust obscuration episode of the donor star, characterized by the formation of large dust grains (reaching sizes of $a\approx 2.0\,\mu$m) and increased mass loss (reaching $\dot{M} \approx 2.1 \times 10^{-5}$\,M$_\odot$\,yr$^{-1}$).
    \item The temporal coincidence of the enhanced mass loss with the X-ray flare suggests a causal connection, and we show that the observed timescales and peak-luminosity of the X-ray outburst is consistent with a wide-orbit ($\sim 50$\,AU) compact object undergoing spherical Bondi accretion from the enhanced mass loss episode. 
\end{itemize}

While the known SyXRBs have been suggested to be either undergoing Roche lobe overflow (GX\,1+4) or steady wind-accretion (e.g. 4U\,1954+31, 4U\,1700+24), \srgt\ presents a unique case of a system where the orbit is wide enough such that the steady wind-accretion results in only very faint X-ray emission. The intense mass loss episode associated with dust formation episode makes it conspicuous in the X-ray sky, and presents the first unambiguous identification of the long-speculated connection between donor mass loss and X-ray activity in SyXRBs. Interestingly, the carbon star companion to the SyXRB CXOGBS\,J173620.2-293338 also shows evidence for dust obscuration episodes in its light curve \citep{Hynes2014}; however there are no constraints on the long term X-ray behavior of the source. \srgt\ demonstrates the unique potential of combining long-term infrared and X-ray time domain surveys in revealing a holistic picture of X-ray outbursts in SyXRBs, which are now being revealed as the dominant contribution to the population of faint X-ray transients in the Milky Way \citep{Bahramian2021}.

Similar to \srgt, there is likely a large population of quiescent NSs/BHs with wide-orbit companions that are otherwise unremarkable. As the donor stars evolve to later stages, X-ray activity caused by enhanced mass loss and dust formation may thus be ubiquitous in wide-orbit SyXRBs. In fact, sporadic dust formation episodes are rare among O-rich evolved stars \citep{Bedding2002}, but are known among many symbiotic O-rich Miras \citep{Munari1988, Munari1989, Jurkic2012a}. Given the typical large dust grain sizes formed during this period, NIR surveys are ideally suited to detect the dramatic fading episodes. Thus, combining deep hard X-ray surveys like ART-XC with long-term optical/NIR monitoring from deep surveys like WINTER \citep{Lourie2020} and the {\it Rubin} observatory \citep{Ivezic2019}, it may soon be possible to `predict' the onset of accretion outbursts of all Galactic symbiotics by searching for dust forming bright IR sources coincident with X-ray counterparts.\\

\facility{{\it ROSAT}, {\it INTEGRAL}, {\it XMM Newton}, SRG: ART-XC, {\it NICER}, {\it Swift}, {\it NuSTAR}, PO: 0.3m (PGIR), PO: 1.2m (ZTF), PO: 1.5m (SED Machine), PO: Hale (DBSP, TripleSpec), PanSTARRS, NEOWISE, IRTF (SpeX), Magellan: Baade (FIRE), SSO: ATT (WiFeS)}   

\software{\texttt{spextool} \citep{Cushing2004}, \texttt{xtellcor} \citep{Vacca2003}, \texttt{pypeit} \citep{pypeit}, \texttt{nustardas} (V2.1), Stingray (V0.3) \citep{Bachetti2021}, \texttt{xspec} \citep{Arnaud1996}, \texttt{emcee} \citep{Foreman-Mackey2013}, DUSTY \citep{Ivezic1997}, HEAsoft (v6.29c; HEASARC 2014)}

\section*{Acknowledgements}

We thank the anonymous referee for a careful reading of the manuscript and providing valuable feedback that helped improve its contents. We thank T. Maccarone, D. Rogantini, P. Pradhan, R. Remillard and M Guenther for valuable discussions. We thank the {\it Swift}, {\it NICER} and {\it NuSTAR} teams for approving the Target of Opportunity observations of the transient. Support for this work was provided by NASA through the NASA Hubble Fellowship grant \#HST-HF2-51477.001 awarded by the Space Telescope Science Institute, which is operated by the Association of Universities for Research in Astronomy, Inc., for NASA, under contract NAS5-26555.

PGIR is generously funded by Caltech, Australian National University, the Mt Cuba Foundation, the Heising-Simons Foundation, the Bi-national Science Foundation. PGIR is a collaborative project among Caltech, Australian National University, University of New South Wales, Columbia University, and the Weizmann Institute of Science. This work is based on data from {\it Mikhail Pavlinsky} ART-XC X-ray telescope aboard the SRG observatory. The SRG observatory was built by Roskosmos in the interests of the Russian Academy of Sciences represented by its Space Research Institute (IKI) in the framework of the Russian Federal Space Program, with the participation of the Deutsches Zentrum für Luft-und Raumfahrt (DLR). The SRG spacecraft was designed, built, launched and is operated by the Lavochkin Association and its subcontractors. The science data are downlinked via the Deep Space Network Antennae in Bear Lakes, Ussurijsk, and Baykonur, funded by Roskosmos. The ART-XC team thank the Russian Space Agency, Russian Academy of Sciences and State Corporation Rosatom for the support of the SRG project and ART-XC telescope and the Lavochkin Association (NPOL) with partners for the creation and operation of the SRG spacecraft (Navigator). This paper includes data gathered with the 6.5 meter Magellan Telescopes located at Las Campanas Observatory, Chile. We acknowledge the Visiting Astronomer Facility at the Infrared Telescope Facility, which is operated by the University of Hawaii under Cooperative Agreement no. NNX-08AE38A with the National Aeronautics and Space Administration, Science Mission Directorate, Planetary Astronomy Program. The authors wish to recognize and acknowledge the very significant cultural role and reverence that the summit of Mauna Kea has always had within the indigenous Hawaiian community.

Work of IM, AL and AS was supported by the Russian Science Foundation grant 19-12-00423. RS acknowledges grant number 12073029 from the National Science Foundation of China. CP acknowledges financial support from the Swiss National Science Foundation (SNF).

\appendix

\section{Multi-Epoch Optical/IR Photometry}
\label{sec:app_photo}

\subsection{Ground-based Optical/NIR photometry}

Given the location of the optical/NIR counterpart in a crowded Galactic plane field, we used forced photometry on difference images from ground-based time domain surveys to derive multi-color light curves. For the PGIR data, the reference image creation took place prior to MJD 58800, and hence we used Point Spread Function (PSF) photometry on the PGIR difference images (using the method described in \citealt{De2020c}) for subsequent data. To derive absolute $J$-band fluxes, the difference fluxes were offset to the non-subtracted $J$-band flux estimated from contemporaneous aperture photometry on the science images near the brightness maximum of the source (where the relative flux contamination from nearby sources would be minimum). For data taken prior to reference creation, we used aperture photometry on science images using a $8.7$\arcsec\ aperture. All reported mags are in the 2MASS Vega system.

We retrieved forced photometry at the source location from ATLAS survey data using the public photometry service\footnote{\url{https://fallingstar-data.com/forcedphot/}}. The source is not detected in the $c$ filter, but a variable source is clearly detected in the redder $o$ filter. We derived an $o$-band light curve for the source using difference photometry fluxes offset to the observed flux level in direct photometry near the epochs where the source was brightest. The derived fluxes were converted to AB mag using a zero-point flux of 3631\,Jy.

The source was also observed in survey operations of ZTF. The light curve derived from the public forced photometry \citep{ Masci2019}) service\footnote{\url{http://web.ipac.caltech.edu/staff/fmasci/ztf/forcedphot.pdf}} was found to be noisy due to the very faint optical flux, and hence we derived photometry using a custom pipeline. We retrieved the ZTF $gri$ science images of the field and used archival stacked images from the PanSTARRS survey as template images to carry out image subtraction (as described in \citealt{De2020b}) followed by forced PSF photometry at the source location. The absolute flux level was calibrated using the source optical fluxes in the PS1 stacked catalog and converted to AB mag. The source is not detected in the ZTF $g$ filter, but a variable source is detected in both the $r$ and $i$ filters.

The source was also observed at multiple epochs during the PS1 surveys. For the redder $i$, $z$ and $y$ bands, we accumulated the PS1 single epoch detection photometry from DR2. For the $g$ and $r$-band PS1 data where the source is much fainter, we retrieved the single epoch PS1 image cutouts to perform subtractions followed by forced photometry as in the case of the ZTF images. The PS1 flux measurements are also reported in AB mags. Since the source is slowly evolving, we binned the source flux over periods of $\approx 7$\,days in creating the light curves from all the surveys and the complete ground-based light curve is provided in Table \ref{tab:oirlc}.

\subsection{Mid-Infrared WISE photometry}

IRAS\,18111-2257 is a bright MIR source, and is reported in the WISE catalogs of the primary 4-band Cryo all-sky survey, 3-band Cryo survey as well as the NEOWISE survey post reactivation of the mission \citep{Wright2010, Mainzer2011}. We retrieved single epoch photometry for the source from the AllWISE and NEOWISE reactivation catalog. Each `visit' of the source consists of several flux measurements taken within a few days, and we derived an average magnitude for the visit by taking the median of all good quality measurements (\texttt{qual\_frame}$> 0$, \texttt{qi\_fact}$>0$, \texttt{saa\_sep}$>0$ and \texttt{moon\_masked}$=0$). The standard deviation of the flux measurements per visit is taken as the uncertainty of the flux. 

The source is nominally brighter than the saturation threshold of WISE in the W1 and W2 bands. As explained in the WISE Data Release document\footnote{\url{https://wise2.ipac.caltech.edu/docs/release/neowise/expsup/sec2_1civa.html}}, photometry for saturated sources is derived by fitting the wings of the non-saturated regions of the PSF but has been observed to be biased in the Post-Cryo Phase. We applied the recommended corrections to the W1 and W2 photometry in the NEOWISE data along with propagating the relevant uncertainties. Since the flux bias for saturated sources in the 4-band and 3-band Cryo mission was observed to be small ($< 0.05$\,mag), we did not apply corrections to the AllWISE photometry. However, we conservatively adopt an uncertainty of 15\% and 23\% for the single epoch AllWISE W1 and W2 photometry respectively, as per the estimated uncertainties in the bias correction. The complete WISE light curve of the source after correction of the aforementioned instrumental effects is given in Table \ref{tab:space_photo}.

\end{document}